\documentclass[final,5p,times,twocolumn]{elsarticle}

\usepackage{lineno,hyperref}

\usepackage{amssymb}
\usepackage{amsmath}
\usepackage[usenames,dvipsnames]{xcolor}

\DeclareGraphicsExtensions{.pdf,.jpeg,.png}
\usepackage{capt-of}
\usepackage{soul}
\usepackage[tight,footnotesize]{subfigure}

\usepackage{diagbox}
\usepackage{algpseudocode}

\usepackage{enumitem}
\usepackage{mathtools}
\usepackage{stfloats}

\usepackage{tikz}
\usetikzlibrary{arrows,decorations.pathreplacing,shapes.misc,patterns,shapes,backgrounds,matrix,positioning,chains,mindmap,shadows,petri,decorations,hobby}
\definecolor{airblue}{rgb}{0.36, 0.54, 0.66}
\definecolor{beaublue}{rgb}{0.74, 0.83, 0.9}
\definecolor{battleshipgrey}{rgb}{0.52, 0.52, 0.51}
\definecolor{whitesmoke}{rgb}{0.96, 0.96, 0.96}
\definecolor{silver}{rgb}{0.75, 0.75, 0.75}
\definecolor{platinum}{rgb}{0.9, 0.89, 0.89}
\definecolor{amaranth}{rgb}{0.9, 0.17, 0.31}
\definecolor{asparagus}{rgb}{0.53, 0.66, 0.42}
\definecolor{arsenic}{rgb}{0.23, 0.27, 0.29}
\definecolor{denim}{rgb}{0.08, 0.38, 0.74}
\definecolor{cornellred}{rgb}{0.7, 0.11, 0.11}
\definecolor{airforceblue}{rgb}{0.36, 0.54, 0.66}

\newcommand\CONDITION[2]%
{\begin{tabular}[t]{@{}l@{}l@{}}
		#1&#2
	\end{tabular}%
}
\algdef{SE}[WHILE]{While}{EndWhile}[1]%
{\algorithmicwhile\ \CONDITION{#1}{\ \algorithmicdo}}%
{\algorithmicend\ \algorithmicwhile}
\algdef{SE}[FOR]{For}{EndFor}[1]%
{\algorithmicfor\ \CONDITION{#1}{\ \algorithmicdo}}%
{\algorithmicend\ \algorithmicfor}
\algdef{S}[FOR]{ForAll}[1]%
{\algorithmicforall\ \CONDITION{#1}{\ \algorithmicdo}}
\algdef{SE}[REPEAT]{Repeat}{Until}{\algorithmicrepeat}[1]%
{\algorithmicuntil\ \CONDITION{#1}{}}
\algdef{SE}[IF]{If}{EndIf}[1]%
{\algorithmicif\ \CONDITION{#1}{\ \algorithmicthen}}%
{\algorithmicend\ \algorithmicif}%
\algdef{C}[IF]{IF}{ElsIf}[1]%
{\algorithmicelse\ \algorithmicif\ \CONDITION{#1}{\ \algorithmicthen}}

\algnewcommand{\IfThenElse}[2]{
	\State \algorithmicif\ #1\ \algorithmicthen\ #2}

\algrenewcommand\algorithmicindent{1.0em}%

\journal{Ad Hoc Networks}

\begin{document}

\let\today\relax

\begin{frontmatter}

\title{A Replication Strategy for Mobile Opportunistic Networks based on Utility Clustering}

\author[uoiaddress]{Evangelos Papapetrou\corref{mycorrespondingauthor}}
\cortext[mycorrespondingauthor]{Corresponding author}
\ead{epap@cse.uoi.gr}

\author[uoiaddress]{Aristidis Likas}
\ead{arly@cse.uoi.gr}

\address[uoiaddress]{Department of Computer Science and Engineering, University of Ioannina, 45110 Ioannina, Greece}

\begin{abstract}
Dynamic replication is a wide-spread multi-copy routing approach for efficiently coping with the intermittent connectivity in mobile opportunistic networks. According to it, a node forwards a message replica to an encountered node based on a utility value that captures the latter's fitness for delivering the message to the destination. The popularity of the approach stems from its flexibility to effectively operate in networks with diverse characteristics without requiring special customization. Nonetheless, its drawback is the tendency to produce a high number of replicas that consume limited resources such as energy and storage. To tackle the problem we make the observation that network nodes can be grouped, based on their utility values, into clusters that portray different delivery capabilities. We exploit this finding to transform the basic forwarding strategy, which is to move a packet using nodes of increasing utility, and actually forward it through clusters of increasing delivery capability. The new strategy works in synergy with the basic dynamic replication algorithms and is fully configurable, in the sense that it can be used with virtually any utility function. We also extend our approach to work with two utility functions at the same time, a feature that is especially efficient in mobile networks that exhibit social characteristics. By conducting experiments in a wide set of real-life networks, we empirically show that our method is robust in reducing the overall number of replicas in networks with diverse connectivity characteristics without at the same time hindering delivery efficiency.
\end{abstract}


\end{frontmatter}

\section{Introduction}
Packet replication has been the dominant routing approach for coping with the intermittent and random connectivity in mobile opportunistic networks~\cite{DF,COORDconf,SimBetTS,EBR,PDF,Redundancy,Spray&Wait,UtilSpray,Greedy,wowmom-cnr}, especially in those where nodes exhibit human mobility such as PSNs (Pocket Switched Networks)~\cite{psns}. The idea behind replication is straightforward; more packet copies increase the probability that a node with a replica will encounter the destination and thus deliver the packet. Yet, replication comes at the cost of more transmissions and increased storage requirements. Therefore, it is imperative to control the level of replication and improve the trade-off between delivery efficiency and cost (both energy and storage related). In other words, it is critical to reduce replication without sacrificing delivery efficiency. So far, the proposed multi-copy routing algorithms work towards this direction but follow two different replication approaches; the ``constrained" (or ``spray-based")~\cite{EBR,SimBetTS,Spray&Wait,UtilSpray} and the ``dynamic" one~\cite{DF,COORDconf,rfc6693}. In the first approach, the source node starts with a predetermined number of replicas ($L$). Each node with multiple copies makes autonomous decisions on how to distribute them. Algorithms in this category differentiate in the decision making regarding the distribution of replicas. The advantage of this approach is that it provides an easy way for controlling replication since $L$ is the upper limit of copies in the network. The downside is that selecting the optimal $L$ is not straightforward since the choice depends on the network properties that are not known beforehand. In ``dynamic" replication, the number of replicas is not predetermined. Instead, each node carrying a packet dynamically creates replicas on a contact basis, i.e., according to the network connectivity. This aspect provides algorithms with the capacity to accommodate networks with diverse characteristics. To control replication levels, in the majority of dynamic schemes, a node chooses a subset of its contacts for creating replicas based on the concept of \textit{utility}, i.e., a value that summarizes the fitness (or quality) of a node for delivering and/or forwarding a message.

In this work, we focus on dynamic replication due to its flexibility and versatility in diverse types of networks. Unfortunately, dynamic schemes exhibit an inclination towards over-replication, i.e., create an unnecessary number of replicas~\cite{DF}. The problem is more severe in the subclass of schemes that endorse a simple ``Compare \& Replicate" approach~\cite{Greedy,wowmom-cnr,DF,rfc6693}. There, a node $v$ replicates a packet to an encountered node $u$ only if the latter has a higher utility. Several methods try to improve this strategy by implementing more elaborate criteria, e.g., require the utility of $u$ to exceed a threshold or evaluate in parallel the number of already created replicas~\cite{rfc6693}. Probably the most efficient of those approaches is the Delegation Forwarding (DF) algorithm~\cite{DF} that exploits $v$'s replication history and mandates that $u$'s utility should exceed the highest utility recorded among $v$'s past contacts. The COORD algorithm~\cite{COORDconf} further improves the performance of DF by enabling packet carriers to coordinate their views about the highest recorded utility among packet carriers.   

Thus far all dynamic schemes make replication decisions using some sort of pair-wise utility-based comparison. In other words, the suitability of a node for carrying a packet replica is decided by comparing its utility to a threshold utility value, e.g., the utility of the packet carrier or the maximum utility among packet carriers, etc. The idea is to place replicas to nodes of increasing delivery capability. We argue that this type of decision making brings significant constraints to our capacity to limit replication since a pair-wise comparison only provides a narrow view of a node's fitness.
In other words, finding a node with a better utility does not always guarantee a significantly improved delivery capability and therefore replicating the packet may be pointless. Instead, we believe that it is possible to obtain a more broad view of a node's fitness by examining how its utility value compares to the utilities of the other nodes in the network. To this end, we capitalize on the observation that, in mobile opportunistic networks and especially in those with human mobility, 
nodes can be classified into groups with diverse delivery capabilities~\cite{Yoneki-Crowcroft,SimBetTS,MilanoJ}. Our intuition is that an analysis of the observed utilities in such a network will bring to light \textit{clusters of utility values that correspond to groups of nodes with different delivery capabilities}, provided that the utility function (or utility metric), i.e., the function used for determining the utility value of a node, effectively captures a node's ability to deliver a message. By classifying nodes to the identified clusters of utilities, it is possible to obtain a network-level view of each node's capability for delivering a message. Then, we can use this knowledge to avoid replication to nodes in the same cluster as they possess similar delivery capabilities. Instead, we choose to \textit{replicate a packet to nodes classified in clusters of increasing delivery capability}.

We portrayed the basic principles of \textit{Cluster-based Replication (CbR)}, a method that incarnates our cluster-driven replication strategy in our previous work~\cite{cbr-wowmom}. In this work, we first shed more light in the clustering property of utilities in real-life networks and provide extensive experimental results to validate it (Section~\ref{sec-formulation}). Then, we delineate the CbR method as a mechanism that works with any utility function and in synergy with any of the  most well-known dynamic replication schemes, such as ``Compare \& Replicate", DF and COORD (Section~\ref{sec-druc}). Furthermore:
\begin{itemize}[leftmargin=10pt]
	\item we theoretically analyze the replication performance of a node implementing CbR in large networks and discuss the computational and time complexity (Section~\ref{subsec-thperformance}).
	\item we provide an in-depth experimental evaluation of CbR using an extended set of diverse contact traces from real-life opportunistic networks as well as an enriched collection of utility functions (Section~\ref{sec-performance} and ~\ref{limited-storage-experiment}). The evaluation corroborates the broad implementation scope of CbR.     
	\item we explore and evaluate various techniques for allowing CbR to keep up with the time-evolving nature of mobile opportunistic networks (~\ref{appendix-sec-updating}). 
	\item we propose $C^{2}bR$, an extension of CbR that implements the concept of cluster-based replication when two utility functions are used for assessing the delivery/forwarding efficiency of a node (Section~\ref{sec-cbr2d}). This is typically the case of social-based routing algorithms. Contrary to such existing algorithms, C$^2$bR does not require a pre-configuration that depends on the network. The experimental evaluation confirms that C$^{2}$bR is robust in networks with diverse characteristics and brings significant cost savings compared to state-of-the-art social-based algorithms.	 
\end{itemize} 
In the rest of the paper, we discuss the system model in Section~\ref{sys-model} and review the related literature on replication schemes for opportunistic networks in Section~\ref{sec-lit}. Section~\ref{sec-concl} summarizes our findings.

\section{System Model, Assumptions ans Scope}\label{sys-model}
In this work we model a mobile opportunistic network as a set of nodes with the ability to communicate wirelessly. Each node $v$ experiences connectivity opportunities with any other node $u$ in the form of a \textit{contact} or \textit{encounter} $<\!v,u,t_{s},t_{e}\!>$, i.e., a one-to-one ability to exchange data in the time interval between $t_{s}$ and $t_{e}$. We assume that contacts occur randomly, i.e., both $t_{s}$ and $t_{e}-t_{s}$ are stochastic processes. The properties of the connectivity experienced by a node such as the rate of contacts and the average contact duration, the average time between contacts, etc, may vary over time, i.e., a node experiences a time-evolving connectivity. In general, the entire network can be modeled as a time-evolving graph. Each node has the ability to store, carry and forward packets exchanged between other nodes in the network. For storing packets to be forwarded, a node $v$ has a storage $Buf_{v}$. We consider both the cases where $Buf_{v}$ is limited or unlimited. Since we consider mobile devices, the main limitation is a node's energy. We do not consider any hard limit on the computational power, however the latter is always a point of consideration due to its relation to the energy consumption. The described system model fits a variety of wireless networking paradigms including PSNs (Pocket Switched Networks)~\cite{psns}, Wireless Mobile Sensor Networks~\cite{wmsn-ref}, Device-to-Device proximity services in the context of 4G and 5G networks~\cite{d2d-ref}, etc.    
	
Finally, we assume that each node $v$ is assigned a \textit{utility} value (or simply \textit{utility}) $U_{v}(d)$ that captures the ability or \textit{fitness} of $v$ to forward/deliver a message to $d$. There are various utility functions (or metrics) that are used to determine a node's utility value. Those metrics are constructed based on some feature of a node's connectivity profile such as the contact rate~\cite{EBR,Greedy}, the time elapsed between successive contacts~\cite{UtilSpray,Fresh}, the probability of node meetings~\cite{Prophet}, as well as features based on the social characteristics of nodes~\cite{SimBetTS, Friendship}. Note that typically a utility is \textit{destination dependent}, i.e., it captures the ability of a node $v$ to deliver packets to their destination. However, there are also \textit{destination independent} utilities that capture a node's ability to interact with other nodes and therefore its fitness for acting as a forwarder regardless of the actual destination. In this case, $U_{v}(d)=U_{v}, \forall d$.

\section{Related Work}\label{sec-lit}
The routing protocols proposed for mobile opportunistic networks with human mobility can be broadly categorized in \textit{single-copy} and \textit{multi-copy} ones. As the names suggest, protocols in the first category use only one copy for each packet while in the second category multiple copies of a packet may exist in the network. Multi-copy schemes are superior to single-copy ones in terms of delivery efficiency. This is because the probability of finding the destination is higher when multiple nodes carry the message. Epidemic routing~\cite{Epidemic} is the extreme of the multi-copy approach; every node carrying a packet forwards a copy to every encountered non-carrier node. Apparently, this strategy results in energy depletion and memory starvation at nodes. Therefore, research efforts have focused in reducing the number of replicas without sacrificing the delivery efficiency. One approach is to use a probabilistic scheme~\cite{epidemic-eval,PDF}, i.e., allow a node to probabilistically create/distribute replicas. Besides the difficulty in setting up the suitable replication probability, this approach is also susceptible to degradation of delivery efficiency.

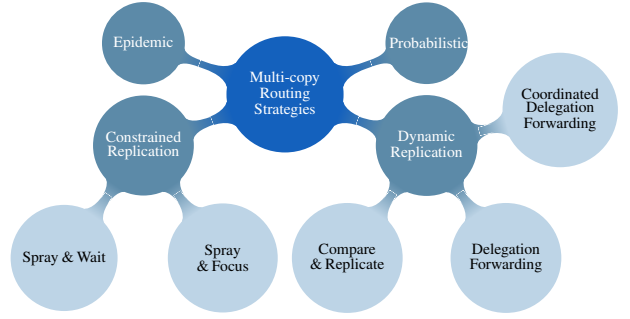
\begin{figure}[!t]
	\centering
	{
		\scalebox{0.65}{
			\begin{tikzpicture}[small mindmap, grow cyclic, concept color=denim, align=flush center,every node/.style={concept}, text=white, align=flush center,text centered,
				level 1/.append style={level distance=3.05cm,sibling angle=90},
				level 2/.append style={level distance=2.8cm,sibling angle=90},
				level 3/.append style={level distance=2.9cm,sibling angle=90},]
				\node[text width=1.5cm, concept color=denim]{{\small  Multi-copy Routing Strategies}}
				child [grow=-160,concept color=airblue,text width=1.5cm,text width=1.8cm] { node {{\small Constrained Replication}}
					child [concept color=beaublue,text=black,grow=-125,text width=2cm] {node {{\small Spray \& Wait}}}
					child [concept color=beaublue,text=black,grow=-55,text width=2cm] {node {{\small Spray\\ \& Focus}}}}
				child [grow=160,concept color=airblue,text width=1.5cm,text width=1.5cm] { node {{\small Epidemic}}}
				child [grow=20,concept color=airblue,text width=1.5cm,text width=1.5cm] { node {{\small Probabilistic}}}
				child [grow=-20,concept color=airblue,text width=1.5cm,text width=1.8cm] { node {{\small Dynamic Replication}}
					child [concept color=beaublue,text=black,grow=-125,text width=2cm] {node {{\small Compare\\ \& Replicate}}}
					child [concept color=beaublue,text=black,grow=-55,text width=2cm] {node {{\small Delegation Forwarding}}}
					child [concept color=beaublue,text=black,grow=15,text width=2cm] {node {{\small Coordinated Delegation Forwarding}}}
				};
			\end{tikzpicture}
		}
	}
	\caption{Classification of multi-copy routing strategies for opportunistic networks with human mobility}
	\vspace{-16pt}
	\label{spread_strategies}
\end{figure}

In the deterministic side, there are two prominent approaches; \textit{``Spray-based"} or \textit{``Constrained replication"} and \textit{``Dynamic replication"} (Fig.~\ref{spread_strategies}). In the first class of algorithms, the source node determines the maximum number of replicas ($L$). Then, the spray process distributes those replicas to other nodes on a contact basis, i.e., every node carrying multiple replicas selects which of its contacts will receive some of them. The selection process is either blind~\cite{Spray&Wait}, i.e., every encountered node is eligible for receiving at least one copy, or based on the candidate carrier's utility. More specifically, assume that node $v$ (with a utility value $U_{v}(d)$ for destination $d$) carries a message $p$ destined to $d$ and encounters node $u$ (with utility value $U_{u}(d)$). Then, $p$ is replicated to $u$~\cite{Spray&Wait,UtilSpray,rfc6693} iff:
\begin{equation}\label{relative_criterion}
	U_{u}(d) > U_{v}(d) + U_{th}
\end{equation}
or
\begin{equation}\label{absolute_criterion}
	U_{u}(d) > U_{th}
\end{equation}  
where $U_{th}$ is a protocol parameter used to secure that the new carrier will contribute a minimum utility improvement (first case) or its utility exceeds a threshold (second case). Another point of differentiation between algorithms in the ``spray-based" category is the spraying method itself, i.e., the decision on how many replicas should an eligible node receive. The most popular strategies are for a node to hand over half of its replicas (binary spray)~\cite{Spray&Wait,UtilSpray} or a fraction of them that depends on $U_{v}(d)$ and $U_{u}(d)$~\cite{EBR,SimBetTS}. When a node ends up with a single copy, it waits until it meets the destination (Spray \& Wait) or uses the utility-based approach to forward the message (Spray and Focus).

The advantage of Spray-based schemes is that it is possible to control the trade-off between delivery efficiency and the degree of replication by determining $L$. Yet, there is an important downside; choosing the optimal $L$ is not trivial since this depends on the network properties. On the other hand, the second multi-copy strategy, known as \textit{``Dynamic replication"}, is more flexible since there is no requirement for predetermining the number of replicas to be created. Instead, every node $v$ carrying a packet follows a utility-based approach and dynamically creates a replica based on the utility of the encountered node $u$. More specifically, in the event of a contact between $v$ and $u$, $v$ implements a \textit{``Compare \& Replicate"} approach~\cite{Greedy,wowmom-cnr,DF,rfc6693}, i.e. forwards a copy to $u$ when (\ref{relative_criterion}) holds with $U_{th}$ set to zero. There are also other, less popular, approaches that relax or enforce (\ref{relative_criterion}) by co-evaluating how many replicas have been created so far or whether $U_{u}(d)$ exceeds a fixed threshold~\cite{rfc6693}. A point of criticism for this approach is that it frequently favors over-replication~\cite{DF}. And this is true regardless of the utility choice, although the latter impacts the algorithm performance. To tackle the problem, Delegation Forwarding (DF)~\cite{DF} introduces a replication strategy that exploits the \emph{history of a node's observations}. To explain, let us consider the case of a contact between $v$, that carries a packet $p$ destined to $d$, and $u$. Then, $p$ is replicated to $u$ iff:
\begin{equation}\label{delegation_criterion}
	U_{u}(d)>\tau^{p}_{v} \; (=\!\max_{k \in N_{v}}\{U_{k}(d)\})
\end{equation}
where $N_{v}$ is the set of all nodes that $v$ has met since the reception of $p$. $\tau^{p}_{v}$ is the delegation threshold that $v$ knowns for $p$, i.e., the highest utility recorded so far among the nodes that received $p$. The idea here is clear; there is no point in replicating a packet to $u$ if another node with a higher utility already has the packet. COORD~\cite{COORDconf} builds on the idea of DF to further reduce replication without impacting delivery efficiency. It makes the observation that $\tau^{p}_{v}$ captures only $v$'s perspective of the highest utility among the packet carriers. Therefore, it enables carrier nodes to coordinate their views, i.e., exchange their thresholds, to obtain a more accurate view of the actual threshold in the network. More specifically, a packet carrier $v$ can take advantage of recurrent contacts with a node $u$ to update its own threshold based on $u$'s threshold. That is, instead of setting its threshold as $\tau^{p}_{v}=\!\max_{k \in N_{v}}\{U_{k}(d)\}$, like in the DF case, it updates its threshold as   
	\begin{equation}\label{coord_update}
		\tau^{p}_{v} \; =\!\max\{\max_{k \in N_{v}}\{U_{k}(d)\},\tau^{p}_{u}\}
	\end{equation}
	In this way, $v$ incorporates $u$'s knowledge into $\tau^{p}_{v}$ and can reduce replication because it has a more accurate view of the highest utility among packet carriers.
Finally, Gao et al.~\cite{Redundancy} also focus on limiting packet redundancy. However, this approach is applicable only to a small class of utility metrics.

All the aforementioned routing protocols share a common ground; they take advantage of a node's contact history to set up the routing strategy. Another very interesting and promising routing approach, applicable to networks that exhibit a social structure, is to exploit not only the node's contact history but also other social information that is available about the network nodes (e.g., Facebook friends or interest of network participants). Algorithms in this class~\cite{ml-sor,peoplerank,mobiclique,sprint}, which is known as \textit{multi-layer social network routing}, exhibit improved routing performance provided that multi-layer social network information is available. There are also available experimental traces that provide multi-layer social information~\cite{unical14,SigComm-dataset}.

Our work focuses on algorithms that exploit only a node's contact history and more specifically on ``dynamic" replication schemes due to their capacity to accommodate networks of diverse characteristics. Our key observation is that, contrary to other schemes that make replication decisions using simple pair-wise comparisons such as in (\ref{relative_criterion}) and (\ref{delegation_criterion}), we can exploit utility information from across the network and optimize the replication process by employing the clustering of observed utility values. We discuss our approach in detail in the following section.

\section{The Clustering Property of Utility values}\label{sec-formulation}

The key concept in ``dynamic" replication schemes is to make replication decisions based on a simple pair-wise comparison involving the individual utilities of the encountering nodes like in (\ref{relative_criterion}) and (\ref{absolute_criterion}). As mentioned, the downside of this strategy is its tendency towards over-replication. Both DF and COORD algorithms target at this drawback by requiring $U_{u}(d)$ to be greater than $\tau^{p}_{v}$, i.e., $v$'s perception of the highest utility among packet carriers (refer to (\ref{delegation_criterion})). Although both algorithms provide an important performance improvement, they do not tackle the root of over-replication which is the limited potential of the ``pair-wise utility comparison" based strategy adopted in (\ref{relative_criterion})-(\ref{delegation_criterion}). A closer look at (\ref{relative_criterion})-(\ref{delegation_criterion}) reveals that the underlying idea is to improve the utility of the carrier as the packet moves towards the destination. The challenge here is to identify what constitutes a suitable minimum utility improvement $\delta U_{min}$, for replicating a packet. Choosing to replicate a packet to candidates that produce a small or minimal $\delta U_{min}$ may result in over-replication. On the other hand, a high $\delta U_{min}$ may result in rejecting the majority or all of the candidate carriers and thus the packet may never reach the destination. The problem is well-known and has been treated by adding $U_{th}$ ($\coloneqq \delta U_{min}$) in (\ref{relative_criterion}).

Yet, determining the optimal $\delta U_{min}$ is a challenge that depends on a series of complex factors such as:
\begin{itemize}[leftmargin=10pt]
	\item the utility function, i.e., how a node's forwarding quality is mapped to a value, since this determines the range of values assigned to candidate carriers. A utility function producing a small value range calls for a small $\delta U_{min}$ and vice versa. 
	\item the network dynamics, i.e., the number and quality of contacts, because they affect the distribution of utility values assigned to nodes. If all nodes share similar connectivity profiles, this results in similar utility values and thus promotes the choice of small $\delta U_{min}$ in order to avoid under-replication. However, this is not necessarily the case if the network consists of nodes with diverse connectivity profiles.
	\item the distance (utility-wise) between the packet carrier and the destination. A large distance may require a small $\delta U_{min}$ to allow the packet to quickly move towards the destination. 
\end{itemize}
Based on the previous discussion, it becomes apparent that using a pair-wise utility comparison approach for making replication decisions is insufficient. 

In this work, we argue that a carrier node $v$, when presented with a forwarding opportunity to a node $u$ with $U_{u}(d)$, instead of just making a local scope pair-wise comparison as in (\ref{relative_criterion})-(\ref{delegation_criterion}), could make a better decision by obtaining a network-wide assessment of $U_{u}(d)$'s importance using the distribution of utility values assigned to other nodes. 
Clearly, it is impossible for a node $v$ to become aware of the aforementioned distribution. Therefore, we opt to use $v$'s perception of this distribution which is the \textit{distribution of utility values} formed by $v$'s past contacts, i.e., the set of values $\{U_{k}(d)\}_{k \in C_{v}}$, where $C_{v}$ is the set of $v$'s past contacts. In this context, the key issue is to determine \textit{how $v$ could exploit the distribution of utility values to identify important replication opportunities}. The answer highly depends on the characteristics of this distribution which in turn depend on the network dynamics. The analysis of contact traces from real networks with human mobility has clearly demonstrated that the nodes of such networks can be classified based on the contact properties into distinct groups~\cite{Yoneki-Crowcroft,MilanoJ}, each one corresponding to a different level of delivery capability. Recall that a utility metric is constructed based on one or more features of a node's contacts. Bearing this in mind, it is reasonable to expect that, for any well-structured utility, \textit{the grouping of nodes will show up as clusters of utility values}. If this is the case then our strategy could \textit{decide} whether a contact $u$ should receive a packet copy \textit{based on the group that $u$ belongs to}, i.e., \textit{instead of making a decision based on $U_{u}(d)$ we decide based on the characteristics of the cluster that $U_{u}(d)$ belongs to}.

\begin{figure*}
	\centering
	\subfigure[]{\includegraphics[width=0.29\linewidth]{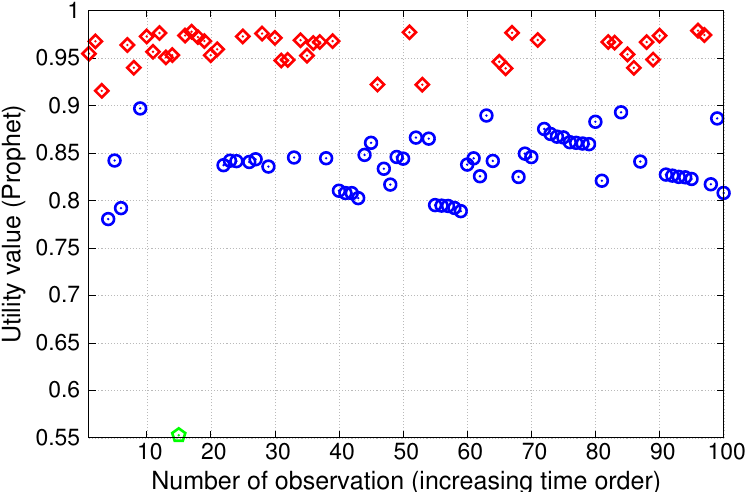}\label{R-Pro}}\hspace{-5pt}
	\subfigure[]{\includegraphics[width=0.29\linewidth]{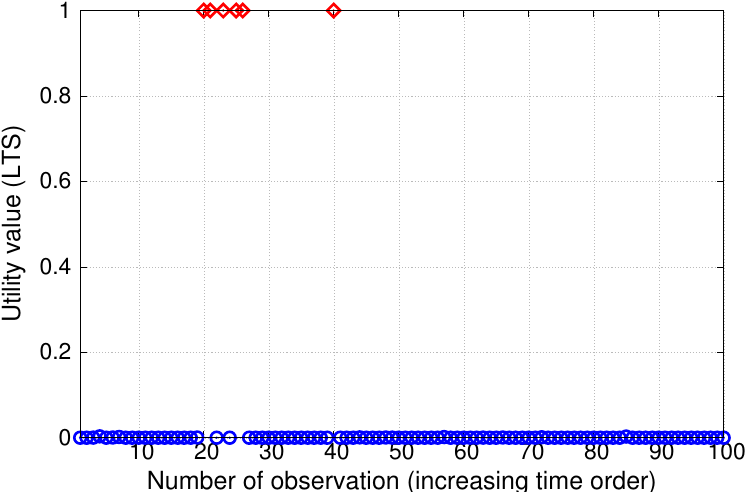}\label{R-LTS}}\hspace{-5pt}	
	\subfigure[]{\includegraphics[width=0.29\linewidth]{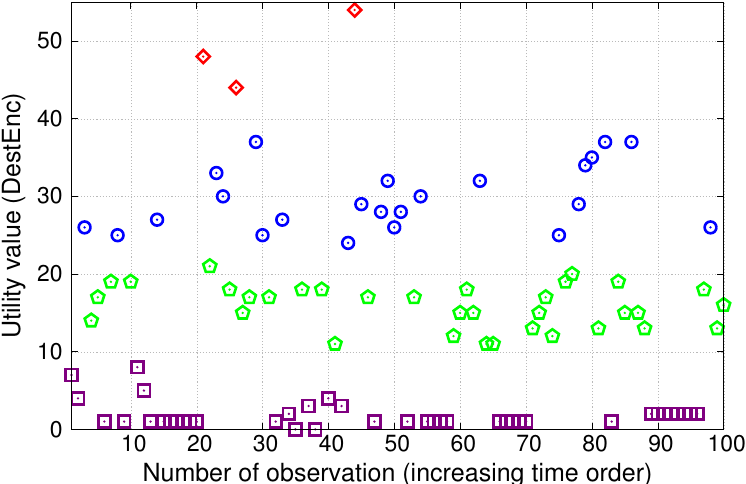}\label{R-Dst}}\hspace{-5pt}%
	\vspace{-12pt}	
	\caption{Clustering of utility values observed by node $v\; (=23)$  for destination $d\; (=50)$, Replication method: ``Compare \& Replicate", Trace: Reality~\cite{Reality-dataset}, Utility function: (a) Prophet~\cite{Prophet}, (b) LTS~\cite{UtilSpray}, and (c) DestEnc~\cite{DF}.}
	\label{validation-res-1}
	\vspace{-12pt}
\end{figure*}
\begin{figure*}
	\centering
	\subfigure[]{\includegraphics[width=0.29\linewidth]{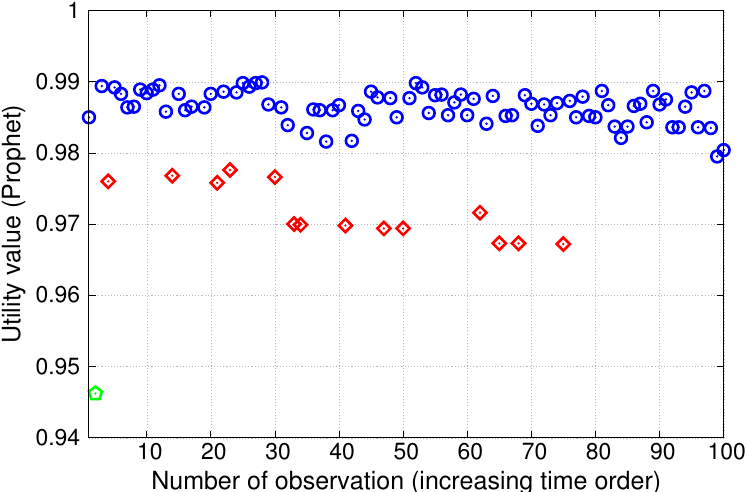}\label{Sig-Pro}}\hspace{-5pt}
	\subfigure[]{\includegraphics[width=0.29\linewidth]{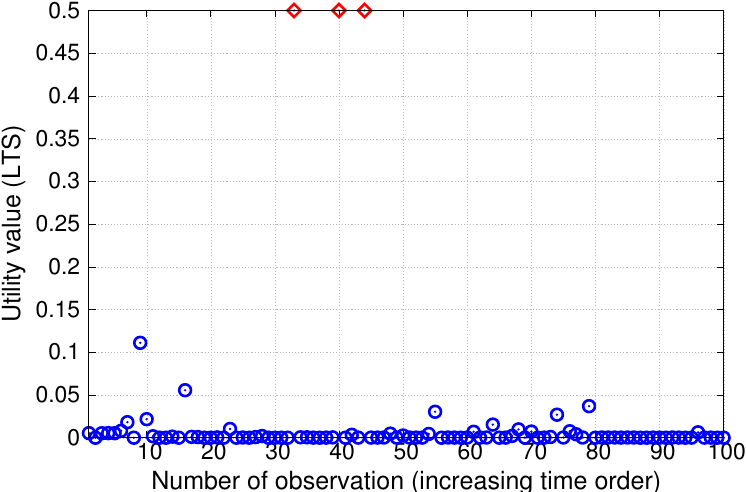}\label{Sig-LTS}}\hspace{-5pt}		
	\subfigure[]{\includegraphics[width=0.29\linewidth]{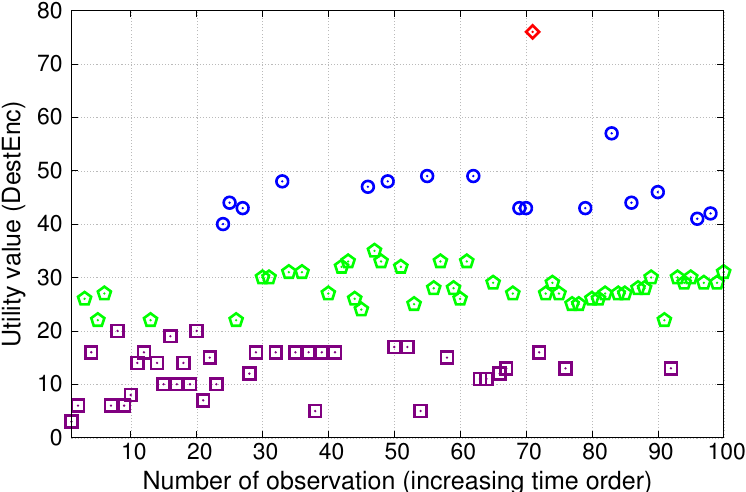}\label{Sig-Dst}}%
	\vspace{-12pt}
	\caption{Clustering of utility values observed by node $v\; (=63)$  for destination $d\; (=31)$, Replication method: ``Compare \& Replicate", Trace: Sigcomm~\cite{SigComm-dataset}, Utility function: (a) Prophet~\cite{Prophet}, (b) LTS~\cite{UtilSpray}, and (c) DestEnc~\cite{DF}.}\label{validation-res-2}
	\vspace{-12pt}
\end{figure*}
To validate the clustering tendency of utility values, we conducted a set of simulation experiments using the ``Compare \& Replicate" approach as the routing algorithm with different utility functions in various real-life contact traces. For every node $v$ we recorded the utilities announced by its contacts for each destination $d$, i.e., the set of values $\{U_{i}(d)\}_{i \in C_{v}}$. After the simulation, we performed an offline calculation to identify clusters of values. To this end, $\forall \;v,d$ pair, we used the $k$-Means clustering algorithm~\cite{kmeans} on the obtained set of one-dimensional data $\{U_{i}(d)\}_{i \in C_{v}}$. We executed $k$-Means for different values of $k$ and the appropriate value, for each set $\{U_{i}(d)\}_{i \in C_{v}}$, was automatically selected using the Silhouette criterion~\cite{silhouette}.
More details about the simulation setup, including the utilities used, can be found in Section~\ref{sec-performance}. Fig.~\ref{validation-res-1} illustrates a series of 100 utility values for the destination with id 50 recorded by the node with id 23 in the Reality trace~\cite{Reality-dataset}. The values are presented in the order they were recorded and different colors (and point types) represent the different clusters of values produced by our approach. The three subfigures correspond to the same experiment but with three different utility functions proposed in the literature, namely Prophet~\cite{Prophet}, LTS~\cite{UtilSpray} and DestEnc~\cite{EBR}. Similar results for the Sigcomm trace ~\cite{SigComm-dataset} are illustrated in Fig.~\ref{validation-res-2}. The grouping of utility values into clusters is evident in all figures. We recorded similar clustering behaviors for utility values from various observer-destination pairs. Since a utility value captures the fitness of a node for forwarding/delivering a packet, we \textit{interpret such clusters of utility values as groups of nodes with different delivery capabilities}. Following this interpretation, \textit{the key idea in our approach is to distribute replicas to nodes that belong to clusters with an increasing delivery capability} in order to avoid unnecessary replications. However, efficiently implementing this strategy highly depends on the observing node and specifically on which cluster the utility of this node belongs to. Therefore, it is imperative to polish the key idea to propose a sophisticated forwarding strategy. We discuss this strategy in detail in Section~\ref{subsec-algorithm}.     

\section{Dynamic Replication with Clusters of Utility}\label{sec-druc}

We call the method that incarnates our cluster-driven replication strategy \textit{Cluster based Replication (CbR)}. CbR is not a standalone algorithm but a mechanism that is integrated into the existing dynamic replication schemes, namely ``Compare and Replicate" (CnR), DF and COORD. Recall that, so far, all those schemes make replication decisions by comparing two utility values. We implement CbR on top of these schemes to transform the decision making process so that, instead of comparing two values, it takes into account the clusters that those values belong to. In the following we will illustrate how CbR works in synergy with the three replication strategies. This will result in three CbR flavors, namely \textit{CbR-CnR}, \textit{CbR-DF} and \textit{CbR-COORD}. CbR consists of three processes:
\begin{itemize}[leftmargin=0.4cm]
	\item \textit{Data Collection and Training}: The training process allows each node to collect a sufficient sample of utility values in order to be able to detect clusters of utility values. During the training period, the node uses the decision making process of the underlying algorithm, i.e., either CnR, DF or COORD. At the end of the training period, the node implements a clustering technique for identifying the utility clusters. Clustering algorithms have been previously used in the context of opportunistic networks but for different purposes, e.g., for identifying node communities based on their contact properties~\cite{BubbleRap} or for fine-tuning the social graph used by social-based algorithms~\cite{knowthy}. They also have been used in a distributed~\cite{TSIROPOULOU} or centralized form~\cite{AMJAD} for efficient data transmission and availability in 5G networks.
	\item \textit{Update}: This process commences after the completion of the training period. Since the network evolves, each node continues to record new utility values through its contacts. These new recordings enrich its view of the distribution of utility values in the network. The update process aims to accordingly refresh the clustering result.
	\item \textit{Decision making}: This is the replication process that exploits the identified clusters of utilities. The process commences after the completion of the training period and operates in parallel with the update process. In contrast to the two other processes, its implementation is different for each of the CbR flavors.
\end{itemize}
In the following subsections we delineate each one of the aforementioned processes.

\subsection{Training and Detecting Clusters of Utility values}\label{subsec-utilitygroups}
CbR starts with a training period, where each node $v$ records the utility values reported by each contact node $u$ for each destination $d$ for which $u$ carries a packet. In other words, $v$ stores, for each destination $d$, a set of values $S_{v}^{d}=\{U_{k}(d)\}_{k \in C_{v}}$, where $C_{v}$ is $v$'s history of contact nodes that carried at least one packet to $d$. In the case of a destination independent metric, i.e., when the reported utility is generic and does not refer to a specific destination, $v$ stores a single set of values $S_{v}=\{U_{k}\}_{k \in C_{v}}$. Note that in all utility-based algorithms, including CnR, DF and COORD, during a contact the two nodes typically exchange their utility values. Therefore, the training process does not involve any additional communication cost. Furthermore, a node $u$ usually reports the utility values on a per packet rather than on a per destination basis, i.e., the utility $U_{u}(d)$ is reported for every packet destined to $d$. Since $U_{u}(d)$ refers to $d$ and not any specific packet, we record it only once during a contact in order to avoid importing noise to the $S_{v}^{d}$ dataset. 
The duration of the training period should allow the collection of a sufficient number of utility samples but at the same time it should not be extremely long in order to facilitate a prompt initiation of the cluster-based replication process. We define the duration of the training period in terms of the number of recorded values. More specifically, the training period ends when $|S_{v}^{d}|\!=\!N_{TR}$, where $N_{TR}$ is a predefined number. Observe that, in the most common case of a destination-dependent utility, the node actually goes through a different training period for every set $S_{v}^{d}$, i.e., for each destination. Moreover, each of these periods may end at a different time because, during a contact, a value is added to $S_{v}^{d}$ only if the contact node carries a packet for $d$.

As mentioned, after the end of the training period a node implements a clustering algorithm on the recorded values. In this work, we choose the \textit{$k$-Means algorithm}~\cite{kmeans} although any clustering algorithm could be used. Our choice is based on the rather simple structure of the clusters observed in the recorded data. This allows us to choose a lightweight algorithm such as $k$-Means since the computational cost is a point of consideration in mobile environments. An important issue in $k$-Means is how to estimate the number of clusters $k$. Recall from Figs.~\ref{validation-res-1} and~\ref{validation-res-2}, that every node may observe a different number of clusters. Therefore, it is not feasible to find a $k$ value that can be used globally. Instead, we follow a more flexible approach where we determine the appropriate number of clusters for each set $S_{v}^{d}$. More specifically, each node executes $k$-Means on $S_{v}^{d}$ for several values of $k$, i.e., $k\!=\!2,3,\ldots,K_{max}$ and obtains $K_{max}-1$ clustering solutions. Next, the quality of each of these solutions is evaluated using the \textit{Silhouette criterion}~\cite{silhouette} and the solution with the highest score is chosen. A pseudocode of the clustering process can be found in~\cite{cbr-wowmom}.

\subsection{Updating the Clustering Result}\label{subsec-updating}

Recall that a utility function relies on a node's connectivity profile, i.e., the average rate and duration of contacts with each node, to assess its forwarding capability and assign a suitable utility value. It is reasonable to assume that in mobile opportunistic networks, especially in those with human mobility, a node's connectivity profile evolves over time, e.g., because a node moves in various locations during different hours of a day. Typically, the time scale of this evolution is relatively large and therefore cannot be captured by the training period which is a one-time process and should be of relatively small duration to timely initiate replication decisions. Hence, we introduce a process that is able to capture changes occurring over relatively long periods of time and update accordingly the clustering structure. This process runs in parallel to the replication one and does not interfere with it.    

Our experimental results indicate that in most cases the clusters of utility values do evolve over time. However, the changes frequently involve the structure and center of the observed clusters rather than their number. Based on this observation, we opted to employ a low-complexity, yet efficient, method for updating the clusters found during the training period. This is the Learning Vector Quantization (LVQ) clustering algorithm~\cite{LVQ} which can be considered as an on-line version of the $k$-Means algorithm. Each time a node records a new utility value $U_{new}$, LVQ decides on which cluster $i$ this value is assigned and subsequently moves the center $c_{i}$ of this cluster towards $U_{new}$, i.e.,
\begin{equation}\label{lvq_update}
	c_{i}^{new} = c_{i} + \alpha (U_{new}-c_{i}) 
\end{equation}
where $\alpha$ is a constant known as the learning rate. In ~\ref{appendix-sec-updating} we explore a set of alternative updating methods and evaluate their impact on CbR's performance.

\subsection{Utilizing Clusters on Replication Decision Making}\label{subsec-algorithm}

After completing the training period, a node is able to use the identified clusters to make replication decisions. In a nutshell, the basic idea of CbR dictates that a node $v$ replicates a packet to $u$ provided that the utility of the latter belongs to a cluster of higher utility values. To implement this simple rule, a node should first rank the identified clusters. This can be easily accomplished since the clustered data are one-dimensional. Thus, we rank the clusters in decreasing order based on their center value, i.e., the cluster with the highest valued center is ranked first. Accordingly, each node $v$ is assigned the rank of the cluster on which its utility value belongs to. In the following, we denote the rank of node $v$ with $R_{v}$. Based on the ranking method, the previous forwarding rule now reads: ``\textit{$u$ receives a packet replica if its utility belongs to a cluster of a higher rank}". Note that this rule is rather stringent and in certain occasions may result in under-replication and thus poor delivery rates. We have identified two occasions where this may occur. The first case is when $U_{v}(d)$ (CnR) or $\tau^{p}_{v}$ (DF or COORD) belongs to the top level cluster of values, i.e., either the carrier $v$ belongs to the top level group of nodes (CnR) or there is a node among packet carriers that belongs to the top level group of nodes. In this case, if $v$ is the packet source the previous rule actually prohibits any replication while if $v$ is an intermediate node the rule blocks any replication within the group of most capable nodes. The second case of potential under-replication occurs when the utility used by $v$ resides in a populous cluster of values and the clusters with a better rank are sparsely populated. In this case, the opportunities for replicating the packet to a better ranked cluster are rare therefore the most probable scenario is that packet replication will involve a substantial delay. The best strategy for both the aforementioned cases is to relax the requirement of replicating the packet to a higher ranked cluster. In other words, it is important to also \textit{allow replication to a node $u$ with a utility in the same cluster provided that $u$'s utility is higher than the utility used by $v$ (traditional decision making).}

\begin{figure}[t]
	\centering
	\subfigure[]{
		\scalebox{0.9}{
			\noindent\fbox{%
				\begin{minipage}{\dimexpr\linewidth-2\fboxsep-2\fboxrule\relax}
					\begin{algorithmic}[1]
						\small
						\Procedure{CbR-CnR} {packet $p$, $U_{v}(d)$, $U_{u}(d)$}
						\IfThenElse {$p \in Buf_{u}$} {exit}
						\State $R_{v} \gets cRank\_of(U_{v}(d))$, $R_{u} \gets cRank\_of(U_{u}(d))$
						\If{$R_{u}\!<\!R_{v}$ \textbf{or} $\big(R_{u}\!\!=\!\!R_{v}$ \textbf{and} $p.rep\!\!=\!\!false\big)$ }
						\If {$U_{u}(d)\!>\!U_{v}(d)$}
						\State Forward $p$ to node $u$
						\State $p.rep \gets true$
						\EndIf
						\EndIf
						\EndProcedure
					\end{algorithmic}
				\end{minipage}%
			}\label{pseudocodeCbR-CnR}
		}
	}\hspace{-5pt}
	\subfigure[]{
		\scalebox{0.9}{
			\noindent\fbox{%
				\begin{minipage}{\dimexpr\linewidth-2\fboxsep-2\fboxrule\relax}
					\begin{algorithmic}[1]
						\small
						\Procedure{CbR-DF} {packet $p$, $\tau_{v}^{\scriptscriptstyle{p}}$, $U_{v}(d)$, $U_{u}(d)$}
						\State $R_{v} \gets cRank\_of(U_{v}(d))$, $R_{u} \gets cRank\_of(U_{u}(d))$
						\State $R_{t} \gets cRank\_of(\tau_{v}^{\scriptscriptstyle{p}})$
						\IfThenElse {$p \in Buf_{u}$} {exit}
						\If {$R_{u}\!\!<\!\!R_{t}\,\text{\textbf{or}}\,\big(R_{u}\!\!=\!\!R_{t}\,\text{\textbf{and}}\,R_{v}\!\!=\!\!R_{t}\big)$}
						\If {$\tau_{v}^{\scriptscriptstyle{p}} < U_{u}(d)$}
						\State Forward $p$ to node $u$
						\State $\tau_{v}^{\scriptscriptstyle{p}}\gets U_{u}(d)$
						\EndIf
						\EndIf
						\EndProcedure
					\end{algorithmic}
				\end{minipage}%
			}\label{pseudocodeCbR-DF}
		}
	}\hspace{-5pt}		
	\subfigure[]{
		\scalebox{0.9}{
			\noindent\fbox{%
				\begin{minipage}{\dimexpr\linewidth-2\fboxsep-2\fboxrule\relax}
					\begin{algorithmic}[1]
						\small
						\Procedure{{\small CbR-COORD}} {packet $p,\tau_{v}^{\scriptscriptstyle{p}},\tau_{u}^{\scriptscriptstyle{p}},U_{v}(d),U_{u}(d)$}
						\State $R_{v} \gets cRank\_of(U_{v}(d))$, $R_{u} \gets cRank\_of(U_{u}(d))$
						\State $R_{t} \gets cRank\_of(\tau_{v}^{\scriptscriptstyle{p}})$
						\If{$p\in$ $Buf_{u}$ \textbf{and} $\tau_{v}^{\scriptscriptstyle{p}} < \tau_{u}^{\scriptscriptstyle{p}}$}
						\State $\tau_{v}^{\scriptscriptstyle{p}}\gets \tau_{u}^{\scriptscriptstyle{p}}$
						\Else
						\If {$R_{u}\!\!<\!\!R_{t}\,\text{\textbf{or}}\,\big(R_{u}\!\!=\!\!R_{t}\,\text{\textbf{and}}\,R_{v}\!\!=\!\!R_{t}\big)$}
						\If{$\tau_{v}^{\scriptscriptstyle{p}} < U_{u}(d)$}
						\State Forward $p$ to node $u$
						\State $\tau_{v}^{\scriptscriptstyle{p}}\gets U_{u}(d)$
						\EndIf
						\EndIf
						\EndIf
						\EndProcedure
					\end{algorithmic}
				\end{minipage}%
			}\label{pseudocodeCbR-COORD}
		}	
	}%
	\vspace{-10pt}
	\caption{Pseudocode of the replication process for a packet $p$ when the carrier node $v$ encounters $u$: a) CbR-CnR, b) CbR-DF and c) CbR-COORD.}\label{pseudocodeCbR-Replication}
	\vspace{-18pt}
\end{figure}
Fig.~\ref{pseudocodeCbR-Replication} presents the pseudocode of CbR when implemented on top of CnR (Fig.~\ref{pseudocodeCbR-CnR}), DF (Fig.~\ref{pseudocodeCbR-DF}) and COORD (Fig.~\ref{pseudocodeCbR-COORD}). The pseudocode is executed for a packet $p$ when the carrier node $v$ encounters node $u$. Note that, in the case of CbR-CnR, the pseudocode is actually the same as in CnR with the single addition being line 4. Recall that in CnR replication decisions are made using (\ref{relative_criterion}) which can also be found in line 5 of the pseudocode. Line 4 realizes our cluster based approach by introducing the requirement $R_{u}\!\!<\!\!R_{v}$, where $R_{v}$ and $R_{u}$ are the ranks of $v$ and $u$ respectively. Both $R_{v}$ and $R_{u}$ can easily be retrieved by simply checking the proximity of $v$'s and $u$'s utility value to the centers of the clusters (procedure $cRank\_of(\cdot)$). We mitigate the risk of under-replication (both identified cases) by moving from the basic criterion $R_{u}\!\!<\!\!R_{v}$ to a relaxed one ($R_{u}\!\!=\!\!R_{v}$) if $v$ has not yet replicated $p$. We distinguish non-replicated packets from replicated ones using a single bit in the packet's header ($p.rep$). As soon as $p$ is forwarded, $p.rep$ is set to 1 and the relaxation is canceled. Note that, in contrary to the $R_{u}\!\!<\!\!R_{v}$ case, it is possible that $U_{u}(d)\!<\!U_{v}(d)$ when $R_{u}\!\!=\!\!R_{v}$. Therefore, the original CnR rule (line 5) is used to control replication in such cases. 

We follow a similar approach when integrating CbR into DF and COORD. Recall that in both DF and COORD, when the packet carrier $v$ encounter $u$, the replication decision is made using (\ref{delegation_criterion}), where $\tau_{v}^{p}$ is $v$'s perception of the highest utility among the carriers of $p$. The two algorithms only differ in the way that $\tau_{v}^{p}$ is updated. Since the decision making criterion is common in DF and COORD, the implementation of the CbR rule is the same in CbR-DF (line 6, Fig.~\ref{pseudocodeCbR-DF}) and CbR-COORD (line 7, Fig.~\ref{pseudocodeCbR-COORD}). Again, the pseudocode of CbR-DF (CbR-COORD) is the same as in DF (COORD), with the only difference being the addition of line 5 (line 7). Regarding the CbR replication rule, observe that the original rule $\tau_{v}^{p}\!<\!U_{u}(v)$ transforms into $R_{u}\!<\!R_{t}$, where $R_{t}$ is the rank of the cluster that $\tau_{v}^{p}$ belongs to. Here, we follow a more efficient approach for relaxing this rule and avoid the two cases of under-replication. More specifically, we allow replication when $R_{u}\!=\!R_{t}$ provided that $R_{v}\!=\!R_{t}$. The latter equality means that $U_{v}(d)$ and $\tau_{v}^{p}$ reside in the same cluster. In other words, the packet carrier $v$ and the carrier with the highest utility have similar delivery capacity, i.e., the packet has not moved to a better cluster. When this happens $\tau_{v}^{p}$ will be updated to a new value so that $R_{t}\!>\!R_{v}$, which will deactivate the relaxation. Again, when $R_{u}\!=\!R_{t}$ the traditional rule (line 7 in Fig.~\ref{pseudocodeCbR-DF} and line 8 in Fig.~\ref{pseudocodeCbR-COORD}) acts as a safeguard. As a final note, all presented implementations are also compatible with destination independent utility functions.

In~\ref{subsec-thperformance}, we examine the performance of CbR in large networks with many nodes and show that the number of replications performed by a node implementing CbR is $O(k)$, i.e., it depends on the number of detected clusters $k$ and not the network size. Furthermore, we discuss the low computational and communication cost required for implementing CbR. We show that CbR imvolves no additional communication cost. At the same time, the computational complexity of obtaining the utility clusters through the $k$-Means algorithm is only $O(N_{TR}k)$ while  the complexity for updating the clustering result through LVQ is only $O(k)$.

\section{Evaluation}\label{sec-performance}

We evaluate the performance of all CbR flavors under various opportunistic environments. To this end, we use the Adyton~\cite{Adyton} simulator. Adyton includes a plethora of routing protocols and is capable of processing a multitude of well-known contact traces from real-world networks~\cite{crawdad-site}.
\begin{table}[b]
	\vspace{-14pt}
	\caption{Properties of real-world opportunistic traces}
	\centering
	\small
	\begin{tabular}{|l||c|c|c|}
		\hline
		Trace Name & \# Nodes & Duration (days) & Area\\
		\hline \hline
		Infocom '05\cite{haggle-dataset} & 41 & 3 & conference \\
		Sigcomm '09\cite{SigComm-dataset} & 76 & 3.7 & conference \\
		MIT Reality\cite{Reality-dataset} & 97 & 283 & campus \\
		Milano pmtr\cite{pmtr-dataset} & 44 & 18.9 & campus \\
		NCCU\cite{nccu} & 115 & 15 & campus \\
		Cambridge upmc\cite{Cambridge-dataset} & 52 & 11.4 & city \\
		\hline
	\end{tabular}
	\label{Traces}
\end{table}
For the evaluation we use traces that represent opportunistic networks of different scale. More specifically, we used two conference traces, namely Infocom'05~\cite{haggle-dataset} and  Sigcomm'09~\cite{SigComm-dataset}. Additionally, we selected two traces from campuses where the participants, students or students and faculty members, move in a larger area. More specifically, we used the well-known MIT Reality~\cite{Reality-dataset}, the Milano pmtr~\cite{MilanoJ,pmtr-dataset} and the NCCU~\cite{nccu} traces. For the NCCU case, we considered the contact recordings during the entirety of the 15 days as in~\cite{nccu-more}.
Finally, we used the Cambridge upmc dataset~\cite{Cambridge-dataset} which is a city-level trace collected in Cambridge, UK. Table~\ref{Traces} summarizes the characteristics of the selected traces.

Similar to CnR, DF and COORD, CbR is able to work with virtually any utility metric proposed in the literature. Clearly, the choice of utility metric impacts performance and therefore the gains of CbR. Thus, to assess the performance of CbR, we use a collection of six well-known utilities, both destination dependent and independent, that have been proposed in the context of opportunistic routing algorithms. More specifically, we use the following utility metrics:
\begin{itemize}[leftmargin=0.4cm]
	\item DestEnc~\cite{DF}: This utility metric captures the total number of contacts with a specific node. Thus, it is a destination dependent utility metric.
	\item Enc~\cite{Greedy,EBR}: This is the destination independent version of DestEnc. The metric captures the total number of contacts with all network nodes.
	\item LTS~\cite{Fresh,UtilSpray}: This is a destination dependent utility metric receiving values in $[0,1]$. It is inversely proportional to the time elapsed since the last contact with the destination.
	\item Prophet~\cite{Prophet,rfc6693}: It is a destination dependent metric proposed in the context of the well-known PRoPHET algorithm. The metric has the transitive property, i.e. it captures the fitness of a node to deliver a message to its destination not only directly but also indirectly.
	\item SPM~\cite{Friendship}: Social Pressure Metric is destination dependent and captures the friendship between network nodes. It depends on the frequency, the longevity and the regularity of past node contacts.
	\item LastContact~\cite{DF}: This is a destination independent metric expressed as $1/(1+T_{L})$ where $T_{L}$ is the time since the node's last contact with any of the network nodes.
\end{itemize}

Regarding the clustering settings, the analysis of data from real contact traces revealed that using a small value for $K_{max}$ such as 4 is sufficient for capturing reasonable estimates of the number of clusters. Furthermore, we used a training period of 50 samples, i.e., $N_{TR}=50$. After extensive experimentation, we found that there are no significant performance improvements for greater values. Finally, the LVQ learning rate $\alpha$ was set equal to 0.05, i.e., the distance between a newly added value and the center of its cluster is reduced by 5$\%$ by moving the center towards the new value.

In each experiment we use a traffic load of 5000 packets. We choose randomly the source/destination pair for each packet while its generation time is chosen with uniform probability in the interval during which both the source and the destination are present in the network. Each packet is assigned a TTL equal to the 20\% of the trace duration. To eliminate statistical bias and monitor the network in its steady state, we use a warm-up and a cool-down period during which packets are not generated. The duration of both periods is 20$\%$ of the total trace duration. We report the average values of 20 repetitions.

\subsection{Results}\label{subsec-res}

\begin{figure*}
	\centering
	\addtolength{\subfigcapskip}{-4pt}
	\subfigure[]{\includegraphics[width=0.329\linewidth]{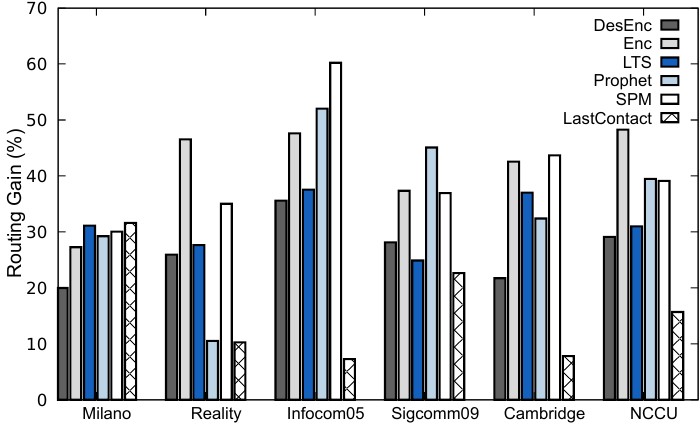}\label{CnRKMeansOH}}
	\subfigure[]{\includegraphics[width=0.329\linewidth]{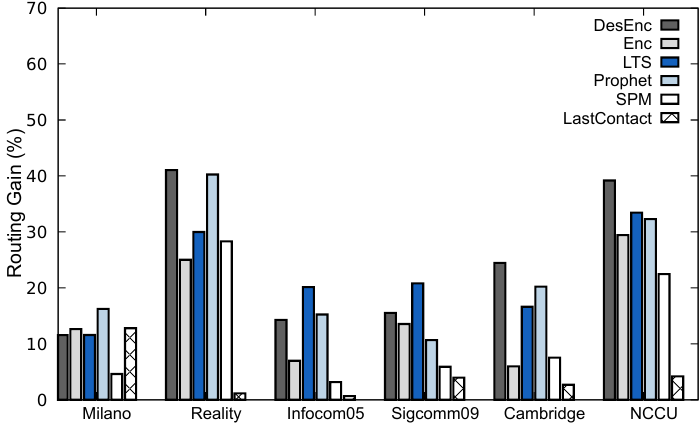}\label{DFKMeansOH}}	
	\subfigure[]{\includegraphics[width=0.329\linewidth]{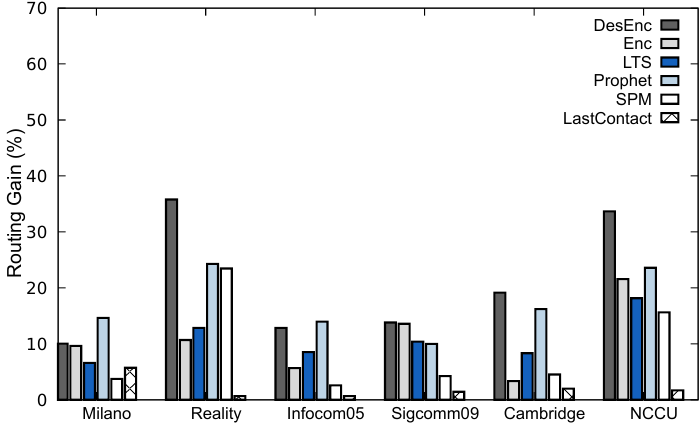}\label{COORDKMeansOH}}%
	\vspace{-12pt}
	\caption{Routing gain of the CbR approach in various traces and for various utility metrics: a) CbR-CnR, b) CbR-DF, and c) CbR-COORD}
	\vspace{-6pt}
	\label{OH-results}
\end{figure*}
\begin{figure*}
	\centering
	\begin{minipage}{0.475\linewidth}
		\includegraphics[width=0.99\columnwidth]{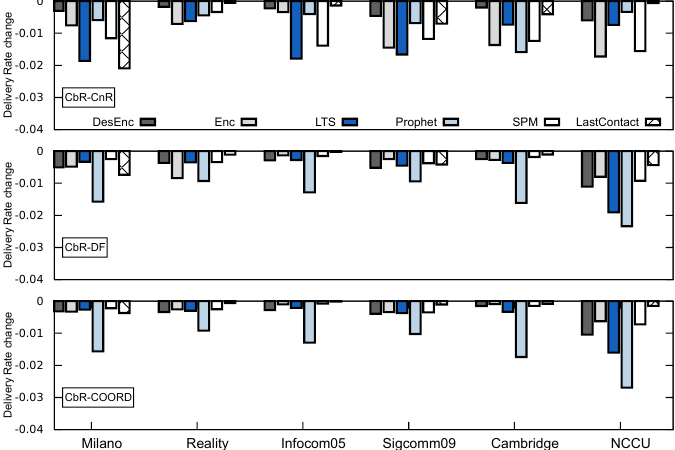}
		\vspace{-8pt}
		\caption{Delivery rate difference of CbR flavors compared to the underlying algorithm in various traces and for various utility metrics}
		\label{DR-results}
	\end{minipage}
	\begin{minipage}{0.05\linewidth}
	\end{minipage}
	\begin{minipage}{0.475\linewidth}
		\includegraphics[width=0.99\columnwidth]{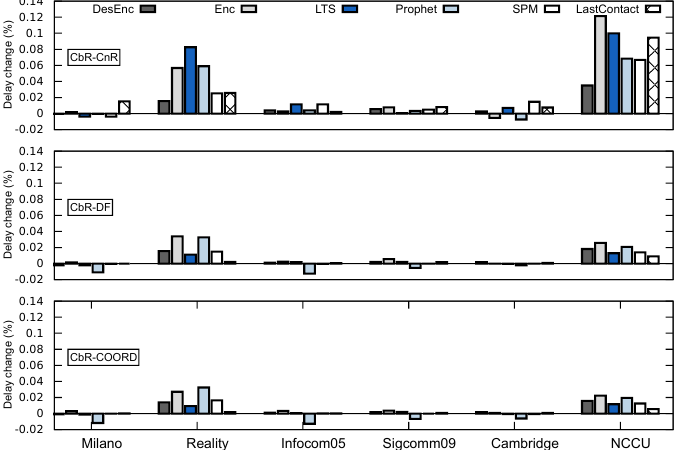}
		\vspace{-8pt}
		\caption{Delivery delay difference of CbR flavors compared to the underlying algorithm in various traces and for various utility metrics}
		\label{Delay-results}
	\end{minipage}
\end{figure*}

In the first experiment we test the performance of all three flavors of CbR in all traces and using each time a different utility metric. To eliminate other interfering factors, we first assume an infinite buffer in each node and a FIFO queueing policy. We use the \textit{routing gain (RG)}, i.e., the percentage of transmissions saved when using CbR, to capture the extend at which CbR reduces the replicas and therefore the number of transmissions. More specifically, we monitor the quantity $(1\!\!-\!\!T_{CbR}/T)\, \%$, where $T$ is the number of transmissions per delivered packet for the underlying algorithm, i.e., either CnR, DF or COORD, and $T_{CbR}$ is the number of transmissions per delivered packet for the CbR flavor of this algorithm. Fig~\ref{OH-results} illustrates the routing gain provided by the CbR approach when used on top of CnR (fig.~\ref{CnRKMeansOH}), DF (fig.~\ref{DFKMeansOH}) and COORD (fig.~\ref{COORDKMeansOH}). In all cases there is a significant gain that, depending on the baseline algorithm and the utility metric, reaches up to an impressive $\sim\!\!60\%$. Reasonably, the routing gain is smaller when CbR is integrated into DF and COORD since those algorithms significantly reduce transmissions on their own. Therefore a smaller room for improvement exists. Still, CbR achieves significant gains of $\sim\!40\%\!\!-\!\!45\%$.

What is of great importance is that CbR's routing gain comes at limited or virtually no cost in delivery. In other words, CbR clearly improves the delivery efficiency-cost trade-off. Fig.~\ref{DR-results} presents the \textit{delivery rate change}, i.e., the quantity $(D_{CbR}/D-1)$ where $D$ is the delivery rate of the baseline algorithm and $D_{CbR}$ is the delivery rate of its CbR version, for all CbR flavors and for all combinations of traces and utility functions. The performance of all CbR flavors is in most cases within $\sim$1$\%$ of the performance of the baseline algorithm and in all cases within $\sim$3.3$\%$. Besides being minimal, this performance degradation can be justified if we bear in mind that even random contacts help nodes communicate. However, such random contacts are not predictable and the only way to exploit them is to increase replication. Furthermore, we show in another experiment (~\ref{limited-storage-experiment}) that, in networks where the storage of nodes is limited, this minimal degradation is almost eliminated and, in many cases, turns into an improvement.  

The reduced level of replication in CbR, as expected, also interferes with the delivery delay. Fig.~\ref{Delay-results} presents the \textit{delay change} (in analogy to the delivery rate change) of CbR flavors. In the cases of the Reality and NCCU traces there is a limited delay increase. This increase is more notable when CnR is the baseline algorithm. However, even in this case, it can be considered acceptable if we keep in mind that CnR achieves low delay because of the excessive replication rate. An easy way to explain those findings is to visualize replication as a process that delivers multiple copies to a destination through different paths. Reducing replication is equivalent to pruning some paths. This delays the packet delivery unless none of the pruned paths is the shortest one in terms of delay, which is rather unlikely. To increase the probability that the shortest delay path will survive pruning, one should assign a high rank to the contacts that this path consists of. However, this responsibility lies with the utility metric and not the replication mechanism. Indeed, note that the delay increase is smaller when the utility metric takes into account connectivity aspects that are related to delay such as the frequency and the regularity of contacts (e.g. the SPM utility). In all other traces, the impact of replication on the delay is negligible. An apparent reason is that all those traces are far more dense, i.e., the contact rate is higher. Therefore, denying a replication opportunity results in a smaller delay increase. Note that in some cases the delay of CbR in fact reduces. This decrease, which is minimal and statistically not important, is attributed to the statistical bias due to the lower delivery rate. 

\section{Two-Dimensional CbR}\label{sec-cbr2d}
\subsection{Making replication decisions using clusters of two utilities}
Up to this point we presented and evaluated CbR with a single utility. Nonetheless, there exist routing algorithms that use two utility functions for making forwarding decisions. This approach typically appears in routing for PSNs (Pocket Switched Networks)~\cite{psns} due to their social structure. Probably the most typical example of social-based routing that capitalizes on two utilities can be found in the SimBet algorithm~\cite{SimBet}. The algorithm adopts two utilities known from social graph analysis, namely ``betweenness"~\cite{bet-1,bet-2} and ``similarity"~\cite{similarity}. Then, it combines them using a normalized weighted sum to form a single utility function, known as ``simbet", based on which forwarding decisions are made.

A first straightforward approach for implementing CbR with the ``simbet" utility is to use the same method as in the single utility case, i.e., apply clustering on the recorded ``simbet" values and then implement one of the algorithms proposed in Section~\ref{sec-druc}. 
In general, we expect this approach to provide some performance gains because the basic idea is the same as in the single utility case; since ``simbet" utility has been proved to be an effective indicator for good forwarders, identifying clusters of ``simbet" values corresponds to detecting groups of forwarders with different delivery capabilities. However, this one-dimensional approach also bears limitations. Each of the similarity and betweenness metrics is associated with a specific social property; similarity is a predictor of social ties and betweenness an indicator of social significance. Nonetheless, it is not clear what is a valid interpretation of the ``simbet" utility. Therefore, when identifying a cluster of high utility nodes it is not clear how this cluster should be interpreted with respect to its social properties. This limits the ways we can exploit this cluster. Furthermore, it has been documented in the related literature that, instead of using the sum of two social-based utilities like in ``simbet", it is beneficial to utilize them independently and sequentially depending on the social proximity of the packet carrier to the destination~\cite{BubbleRap}.

\begin{figure}
	\centering
	\subfigure[]{
		\scalebox{0.9}{
			\noindent\fbox{%
				\begin{minipage}{\dimexpr\linewidth-2\fboxsep-2\fboxrule\relax}
					\begin{algorithmic}[1]
						\small
						\Procedure{C$^2$bR-CnR} {pkt $p$, $S_{v}(d)$, $S_{u}(d)$, $B_{v}$, $B_{u}$}
						\IfThenElse {$p \in Buf_{u}$} {exit}
						\State $R_{v}^{S} \gets cRank\_of(S_{v}(d))$, $R_{v}^{B} \gets cRank\_of(B_{v})$
						\State $R_{u}^{S} \gets cRank\_of(S_{u}(d))$, $R_{u}^{B} \gets cRank\_of(B_{u})$
						\If{$R_{v}^{S}<>1$}
						\State CbR-CnR($p$,$B_{u}$,$B_{v}$)
						\ElsIf{($R_{v}^{S}=1$ \textbf{and} $R_{u}^{S}=1$)}
						\State CbR-CnR($p$,$B_{u}$,$B_{v}$)
						\EndIf
						\EndProcedure
					\end{algorithmic}
				\end{minipage}%
			}\label{pseudocodeC2bR-CnR}
		}
	}\hspace{-5pt}
	\subfigure[]{
		\scalebox{0.9}{
			\noindent\fbox{%
				\begin{minipage}{\dimexpr\linewidth-2\fboxsep-2\fboxrule\relax}
					\begin{algorithmic}[1]
						\small
						\Procedure{C$^2$bR-DF} {pkt $p$, $\tau_{v}^{\scriptscriptstyle{p},S}$,$\tau_{v}^{\scriptscriptstyle{p},B}$, $S_{v}(d)$,$S_{u}(d)$, $B_{v}$, $B_{u}$}
						\State $R_{v}^{S} \gets cRank\_of(S_{v}(d))$, $R_{v}^{B} \gets cRank\_of(B_{v})$
						\State $R_{u}^{S} \gets cRank\_of(S_{u}(d))$, $R_{u}^{B} \gets cRank\_of(B_{u})$
						\State $R_{t}^{S} \gets cRank\_of(\tau_{v}^{\scriptscriptstyle{p},S})$, $R_{t}^{B} \gets cRank\_of(\tau_{v}^{\scriptscriptstyle{p},B})$
						\IfThenElse {$p \in Buf_{u}$} {exit}
						\If {($R_{t}^{S}=R_{v}^{S}$ \textbf{and} $R_{t}^{S}<>1$)}
						\State CbR-DF($p$, $\tau_{v}^{\scriptscriptstyle{p},B}$, $B_{v}$, $B_{u}$)
						\ElsIf {($R_{v}^{S}=1$ \textbf{and} $R_{u}^{S}=1$)}
						\State CbR-DF($p$, $\tau_{v}^{\scriptscriptstyle{p},B}$, $B_{v}$, $B_{u}$)
						\EndIf
						\IfThenElse {$\tau_{v}^{\scriptscriptstyle{p},S} < S_{u}(d)$} {$\tau_{v}^{\scriptscriptstyle{p},S}\gets S_{u}(d)$}
						\EndProcedure
					\end{algorithmic}
				\end{minipage}%
			}\label{pseudocodeC2bR-DF}
		}	
	}\hspace{-5pt}		
	\subfigure[]{
		\scalebox{0.9}{
			\noindent\fbox{%
				\begin{minipage}{\dimexpr\linewidth-2\fboxsep-2\fboxrule\relax}
					\begin{algorithmic}[1]
						\small
						\Procedure{{\small C$^2$bR-COORD}} {pkt $p$, $\tau_{v}^{\scriptscriptstyle{p,S}}$, $\tau_{v}^{\scriptscriptstyle{p,B}}$, $\tau_{u}^{\scriptscriptstyle{p,S}}$, $\tau_{u}^{\scriptscriptstyle{p,B}}$, $S_{v}(d)$, $S_{u}(d)$, $B_{v}$, $B_{u}$} 
						\State $R_{v}^{S} \gets cRank\_of(S_{v}(d))$, $R_{u}^{S} \gets cRank\_of(S_{u}(d))$
						\State $R_{v}^{B} \gets cRank\_of(B_{v})$, $R_{u}^{B} \gets cRank\_of(B_{u})$
						\State $R_{t}^{S} \gets cRank\_of(\tau_{v}^{\scriptscriptstyle{p,S}})$, $R_{t}^{B} \gets cRank\_of(\tau_{v}^{\scriptscriptstyle{p,B}})$
						\If{$p\in$ $Buf_{u}$}
						\IfThenElse {$\tau_{v}^{\scriptscriptstyle{p},S} < \tau_{u}^{\scriptscriptstyle{p},S}$} {$\tau_{v}^{\scriptscriptstyle{p},S}\gets \tau_{u}^{\scriptscriptstyle{p},S}$}
						\IfThenElse {$\tau_{v}^{\scriptscriptstyle{p},B} < \tau_{u}^{\scriptscriptstyle{p},B}$} {$\tau_{v}^{\scriptscriptstyle{p},B}\gets \tau_{u}^{\scriptscriptstyle{p},B}$}
						\Else
						\If {($R_{t}^{S}=R_{v}^{S}$ \textbf{and} $R_{t}^{S}<>1$)}
						\State CbR-COORD($p$, $\tau_{v}^{\scriptscriptstyle{p},B}$, $\tau_{u}^{\scriptscriptstyle{p},B}$, $B_{v}$, $B_{u}$)
						\ElsIf {($R_{v}^{S}=1$ \textbf{and} $R_{u}^{S}=1$)}
						\State CbR-COORD($p$, $\tau_{v}^{\scriptscriptstyle{p},B}$, $\tau_{u}^{\scriptscriptstyle{p},B}$, $B_{v}$, $B_{u}$)
						\EndIf
						\EndIf
						\EndProcedure
					\end{algorithmic}
				\end{minipage}%
			}\label{pseudocodeC2bR-COORD}
		}
	}%
	\vspace{-10pt}
	\caption{Pseudocode of the replication process for a packet $p$ when the carrier node $v$ encounters $u$: a) C$^2$bR-CnR, b) C$^2$bR-DF and c) C$^2$bR-COORD.}\label{pseudocodeC2bR-Replication}
	\vspace{-18pt}
\end{figure}

The latter observation has been the driving force of our second approach which we call two-dimensional CbR or simply $C^{2}bR$. More specifically, we examine betweenness and similarity independently and identify the corresponding clusters. Since betweenness captures the social importance, clusters of betweenness values correspond to groups of nodes with different social importance. On the other hand, similarity, besides being an indicator of future social ties, also reveals social proximity because it is non-zero when the social proximity to the destination is no more than two hops. Therefore, different clusters of similarity values correspond to nodes with different social proximity to the destination. The key concept in all C$^2$bR flavors is simple and in a nutshell can be expressed as follows: ``\textit{Move the message up to the social hierarchy constructed using betweenness until a message reaches a group of nodes with high social proximity (similarity) to the destination. At this time, continue the same strategy but confine it within this group of nodes}". It is well-known that in networks with social heterogeneity a single utility that provides a ranking of nodes cannot perform efficiently~\cite{BubbleRap}. On the other hand, it is also not possible to rely on a metric that captures social proximity to the destination because the source of the packet may be socially far away~\cite{BubbleRap}. C$^2$bR's strategy combines the two utilities to allow a message to move far from the source when needed (source and destination socially apart) and then move the message towards the destination (by using a destination dependent utility).

The previous strategy materializes in three versions of C$^{2}$bR; one based on CnR (C$^{2}$bR-CnR), another based on DF (C$^{2}$bR-DF) and the third based on COORD (C$^{2}$bR-COORD). The pseudocode of the three algorithms is presented in Fig.~\ref{pseudocodeC2bR-Replication} where $S_{v}(d)$ is the similarity of $v$ for $d$ (a destination dependent metric), $B_{v}$ is the betweenness of $v$ (a destination independent metric), $\tau_{v}^{\scriptscriptstyle{p,S}}$ ($\tau_{v}^{\scriptscriptstyle{p,B}}$) is $v$'s perception of the highest similarity (betweenness) among the carriers of packet $p$, $R_{v}^{S}$ ($R_{v}^{B}$) is the rank of the cluster that $v$'s similarity (betweenness) belongs to and $R_{t}^{S}$ ($R_{t}^{B}$) is the rank of the cluster that $\tau_{v}^{\scriptscriptstyle{p,S}}$ ($\tau_{v}^{\scriptscriptstyle{p,B}}$) belongs to. Observe that, similar to CbR-DF and CbR-COORD, C$^{2}$bR-DF and C$^{2}$bR-COORD only differ in the way they update $\tau_{v}^{\scriptscriptstyle{p,S}}$ and $\tau_{v}^{\scriptscriptstyle{p,B}}$ but share a common forwarding strategy which we summarize in the following. If a packet carrier $v$ (including the source) does not belong to the highest similarity cluster then it uses only betweenness and reverts to the simple CbR algorithm, either CbR-DF or CbR-COORD respectively (lines 6-7 in Fig.~\ref{pseudocodeC2bR-DF} and 9-10 in Fig.~\ref{pseudocodeC2bR-COORD}). This corresponds to an attempt to find more socially important forwarders. However, replication stops once $v$ finds out that the packet has been replicated to a node that belongs to a group of better similarity ($R_{t}^{S}\!<\!R_{v}^{S}$). This is to promote replication to nodes with increasingly higher social proximity to the destination. On the other hand, when $v$ belongs to the group of nodes with the highest similarity, i.e., highest social proximity to destination (lines 8-9 in Fig.~\ref{pseudocodeC2bR-DF} and 11-13 in Fig.~\ref{pseudocodeC2bR-COORD}), again it tries to find a more socially important carrier by reverting to the simple CbR flavor with betweenness as the only utility. However, in this case, replication is confined to nodes in the highest similarity cluster ($R_{u}^{S}\!=\!1$), i.e., we do not allow replication that decreases the social proximity to the destination. 
In analogy, C$^2$bR-CnR falls back to simple CbR-CnR with betweenness as the only utility but confines replication within the group of nodes with the highest proximity to the destination, i.e., highest similarity, when the packet carrier $v$ belongs to that group. This is done by requiring the packet recipient to also belong in this group ($R_{u}^{S}\!=\!1$).

\subsection{Two vs One dimensional CbR}

\begin{figure*}
	\centering
	\addtolength{\subfigcapskip}{-4pt}
	\subfigure[]{\includegraphics[width=0.329\linewidth]{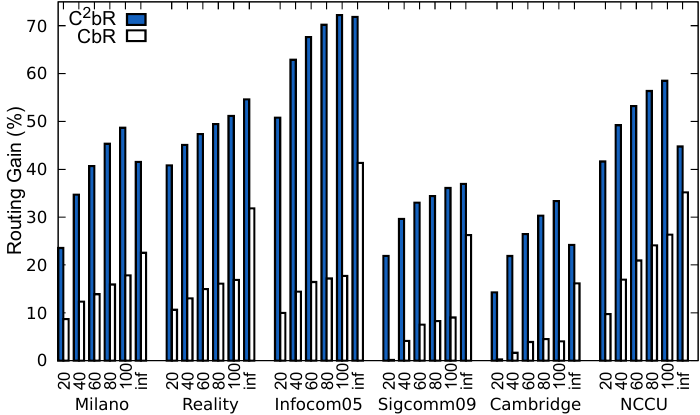}\label{CbR2D-CnR-OH_BUF}}
	\subfigure[]{\includegraphics[width=0.329\linewidth]{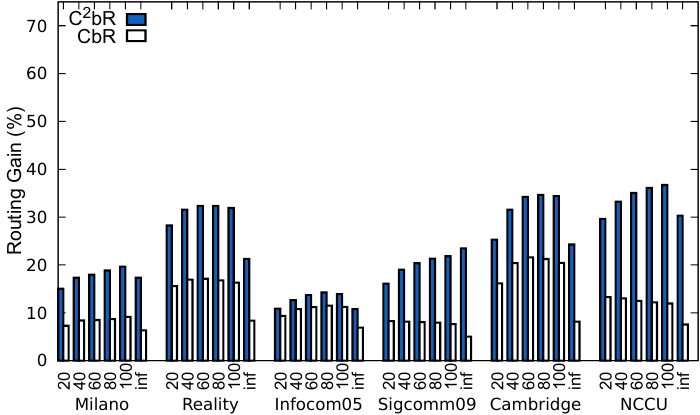}\label{CbR2D-DF-OH_BUF}}	
	\subfigure[]{\includegraphics[width=0.329\linewidth]{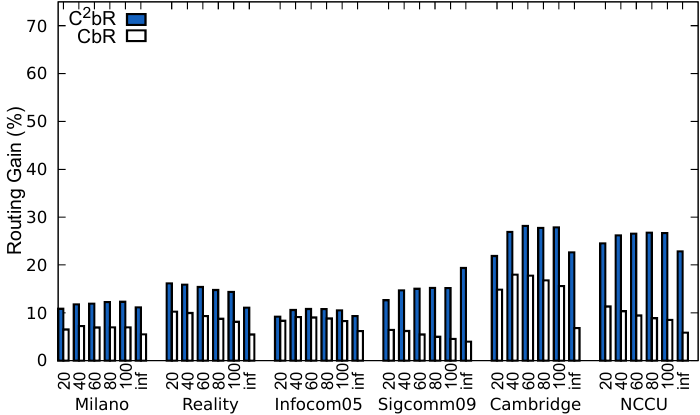}\label{CbR2D-COORD-OH_BUF}}%
	\vspace{-12pt}
	\caption{Routing gain of the C$^{2}$bR and CbR vs nodes' storage capacity for different traces when implemented on top of: a) CnR, b) DF, and c) COORD}
	\vspace{-12pt}
	\label{CbR2D-OH}
\end{figure*}
\begin{figure*}
	\centering
	\addtolength{\subfigcapskip}{-4pt}
	\subfigure[]{\includegraphics[width=0.329\linewidth]{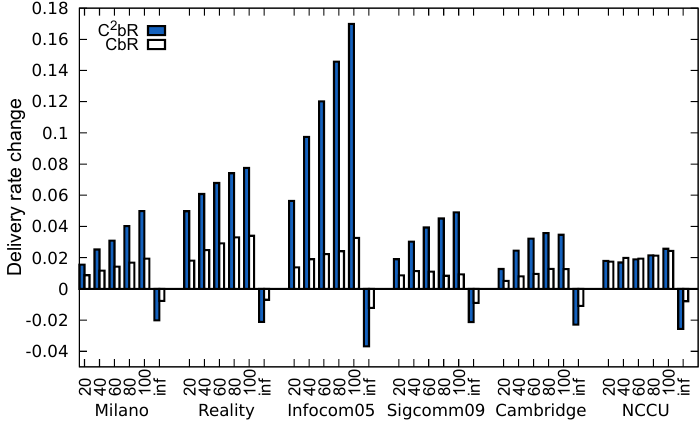}\label{CbR2D-CnR-DR_BUF}}
	\subfigure[]{\includegraphics[width=0.329\linewidth]{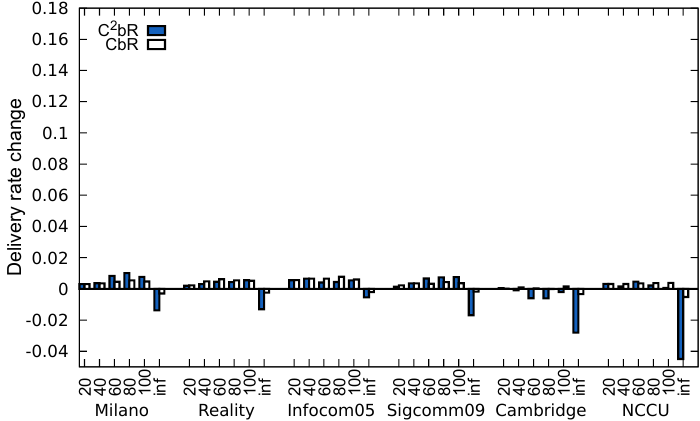}\label{CbR2D-DF-DR_BUF}}		
	\subfigure[]{\includegraphics[width=0.329\linewidth]{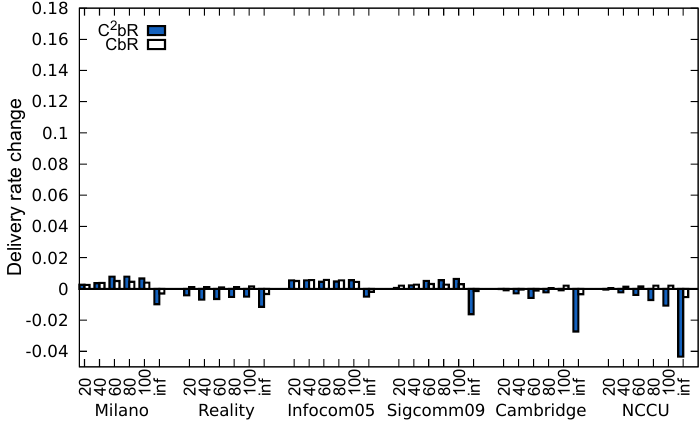}\label{CbR2D-COORD-DR_BUF}}%
	\vspace{-8pt}
	\caption{Delivery rate change of C$^{2}$bR and CbR vs nodes' storage capacity for different traces when implemented on top of: a) CnR, b) DF, and c) COORD}
	\vspace{-16pt}
	\label{CbR2D-DR}
\end{figure*}
To evaluate the performance of C$^2$bR we compare it with the simple CbR approach that uses ``simbet" as the utility function. We test the two approaches on top of CnR, DF and COORD in five different traces and for various node storage capacities. Fig.~\ref{CbR2D-OH} presents the routing gain of both CbR and C$^2$bR with respect to the performance of the underlying algorithm (either CnR, DF or COORD) when it uses ``simbet" as the utility function. As expected, the simple CbR approach provides significant performance improvements in all cases. At the same time, the results confirm our assessment regarding the limitations of CbR and prove that it is possible to achieve vast performance improvements with C$^2$bR. Indeed, in most cases C$^2$bR manages to almost double the routing gain or perform even better. But most impressively, C$^2$bR provides this gain with virtually no trade-off. In fact, C$^2$bR-CnR also significantly improves delivery efficiency (fig.~\ref{CbR2D-DR}) while C$^2$bR-DF and C$^2$bR-COORD achieve virtually the same performance as CbR-DF and CbR-COORD and better performance than the underlying algorithm (i.e., DF and COORD respectively). The only exception is the case of infinite storage capacity where a small performance lag is observed. Again, this is reasonable since both CbR and C$^2$bR significantly limit replication and therefore the probability of exploiting random contacts to deliver messages is smaller. In any case, the observed lag is limited (on average $\sim\!2\%$ and no more than $\sim\!4.5\%$ in the worst case).

\subsection{C$^2$bR as a social-based routing algorithm}

Besides the benefits of C$^2$bR over CbR, we also examine how C$^2$bR compares to social-based routing algorithms. Since C$^2$bR makes routing decisions based only on a node's contact history, we compare it with the most well-known and established social routing algorithms that also rely only on this type of information, namely SimBet~\cite{SimBet} and BubbleRap~\cite{BubbleRap,BubbleRap-conf}. For a comparison that focuses on multi-layer social routing algorithms, i.e., algorithms that exploit other kinds of social information, the interested reader may refer to~\cite{ml-sor}.
SimBet was originally proposed as a single-copy algorithm featuring the ``simbet" utility that we discussed previously. For a fair comparison, we used its follow-up multi-copy version~\cite{SimBetTS}. This falls in the spray-based category of algorithms, i.e., a predetermined number of $L$ packet copies is distributed in the network. The distribution and forwarding of copies depends on the ``simbet" utility of the encountering nodes. In order to produce the ``simbet" utility, we used the proposed weight of 0.5 for both similarity and betweenness, i.e., both have equal importance~\cite{SimBet}. BubbleRap is a multi-copy algorithm that falls in the dynamic replication sub-class~\cite{BubbleRap}. The algorithm requires a community detection mechanism. Its forwarding strategy bears similarities to the one in C$^2$bR. A message is moved up in the global hierarchy, constructed based on the centrality of each node, until it reaches a node in the destination's community. Then, the message is moved within the community using the local hierarchy, constructed based on the local centrality of nodes. Besides the apparent analogy, which is to forward a message up in the hierarchy until it moves in the social vicinity of the destination, C$^2$bR is different from BubbleRap in many aspects. First, C$^2$bR uses ego-betweenness~\cite{egoBet} to approximate betweenness centrality and construct the global hierarchy whereas BubbleRap proposes the use of the average unit-time degree. More importantly, C$^2$bR does not require any community detection mechanism to identify the destination's social neighborhood. Nor it requires any sort of customization that comes with it. Instead, it capitalizes on the metric of similarity to quantify the social proximity to the destination and route the packet in the direction of increasing proximity. Last but not least, C$^2$bR utilizes the concept of cluster-based replication in all phases of forwarding a message towards the destination in order to reduce the incurred cost. To enable distributed community detection by a node in BubbleRap, we implemented the distributed version of the $K$-CLIQUE algorithm discussed in~\cite{BubbleRap} and described in~\cite{BubbleRap-dist-comm}. $K$-CLIQUE requires some customization that depends on the network, namely the value for $K$ as well as a weight threshold for ruling out insignificant, from a social point of view, contacts. We focus our comparison in the Reality trace since it is a typical example of a trace exhibiting social characteristics. Furthermore, this choice allows us to use the parameter values for $K$-CLIQUE that were reported in~\cite{BubbleRap,BubbleRap-conf} for the Reality trace, namely $K=3$ and a threshold of 388800s. This is critical for providing a fair comparison.

\begin{figure}
	\centering
	\addtolength{\subfigcapskip}{-4pt}
	\subfigure[]{\includegraphics[width=0.9\linewidth]{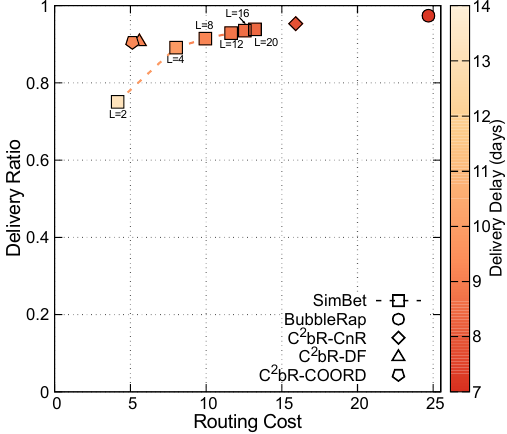}\label{DRvsOHvsDelay-Multi-Cataclysm}}
	\vspace{-8pt}
	\caption{Performance of C$^2$bR schemes compared to SimBet (various values of $L$) and BubbleRap in the Reality trace.}
	\vspace{-18pt}
	\label{C2bR-SimBet-Bubble-0}
\end{figure}
To evaluate the pure replication and forwarding efficiency of all algorithms in terms of the delivery-cost trade-off, we first consider the case of unlimited storage at each node. Moreover, we assume that all copies of a message are instantly deleted upon delivery of this message to the destination. Fig.~\ref{C2bR-SimBet-Bubble-0} presents the performance of all algorithms in terms of delivery ratio, i.e., the percentage of packets successfully delivered to the destination, and routing cost, i.e., the average number of transmissions performed for each message. Furthermore, the performance of each protocol in terms of average delivery delay is presented with different color darkness. C$^2$bR-DF and C$^2$bR-COORD achieve the best delivery-cost trade-off, a confirmation of the effectiveness of the cluster-based approach. SimBet achieves approximately the same delivery efficiency (for $L\!\geq\!8$) at a cost that is $\sim\!2\!-\!2.5$ times greater than the cost of C$^2$bR schemes. Note that obtaining the best performance for SimBet depends on determining the optimal $L$, which is not a straightforward task since it depends on the network. On the other hand, BubbleRap produces an improvement of $\sim\!6\%$ in the delivery ratio but this comes at a high cost ($\sim\!5$ times greater than the cost of C$^2$bR-DF and C$^2$bR-COORD). Moreover, C$^2$bR-CnR achieves a delivery ratio close to BubbleRap (lagging only $\sim\!2\%$) but its cost is only $\sim\!64\%$ of BubbleRap's cost.

It is clear that the previous setting captures the best case performance with respect to the routing cost because it assumes that all copies are immediately deleted upon delivery of a message to the destination. In a real-life setting, it is critical for an algorithm to provide a stopping rule in order to prevent nodes, not meeting the destination, from continuing replication after message delivery. Spray-based schemes impose such a rule since a node left with one copy is not allowed to continue replication. BubbleRap, on the other hand, does not delineate any specific stopping rule. Interestingly, C$^2$bR-DF and C$^2$bR-COORD inherently enforce a stopping policy for each node that in a nutshell can be described as follows: a node stops replication as long as the packet is moved to a node that belongs to a better cluster (see line 6 in Fig.~\ref{pseudocodeC2bR-DF} and line 9 in Fig.~\ref{pseudocodeC2bR-COORD}). These stopping rules are also augmented by using the concept of utility threshold. C$^2$bR-CnR also delineates a stopping policy which is, however, less effective since it does not use the idea of utility threshold. The stopping rule dictates that replication stops when the packet reaches a node in the best cluster (see line 5 in Fig.~\ref{pseudocodeC2bR-CnR}). We extensively experimented with the more realistic scenario where nodes that do not meet the destination erase a packet based on TTL. We found that both SimBet and BubbleRap failed to compete in terms of routing cost with C$^2$bR schemes and especially C$^2$bR-DF and C$^2$bR-COORD. Compared to the previous case (Fig.~\ref{C2bR-SimBet-Bubble-0}), the cost of both C$^2$bR-DF and C$^2$bR-COORD increased slightly, the cost of SimBet increased by up to $30\%$ depending on the value of $L$ and, as expected, BubbleRap's cost escalated dramatically. In other words, C$^2$bR-DF and C$^2$bR-COORD proved to be the most efficient regarding the policy for stopping replication while the strategy of SimBet proved to be inadequate. At the same time BubbleRap's performance collapses due to the lack of any stopping policy.

\begin{figure}[b]
	\vspace{-12pt}
	\centering
	\addtolength{\subfigcapskip}{-4pt}
	\subfigure[]{\includegraphics[width=0.4935\linewidth]{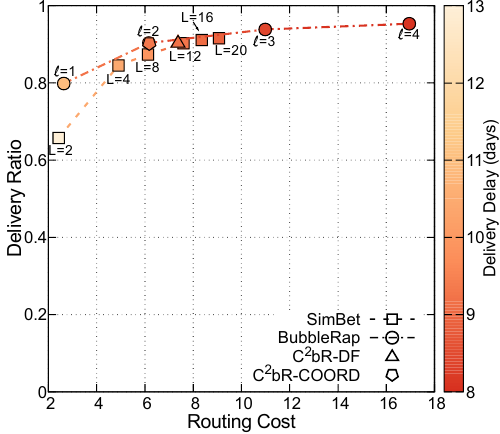}\label{DRvsOHvsDelay-Multi-HL3}}
	\subfigure[]{\includegraphics[width=0.4935\linewidth]{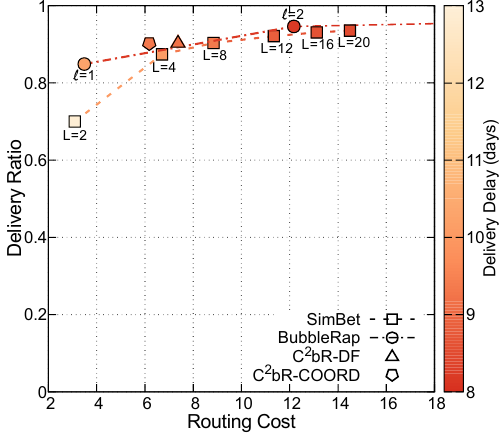}\label{DRvsOHvsDelay-Multi-HL4}}%
	\vspace{-8pt}
	\caption{Performance of C$^2$bR-DF and C$^2$bR-COORD compared to SimBet and BubbleRap for different numbers of replicas. Max hops for SimBet and BubbleRap: a) three (3), b) four (4).}
	\label{C2bR-SimBet-Bubble-1}
\end{figure}
\begin{figure}
	\centering
	\addtolength{\subfigcapskip}{-4pt}
	\subfigure[]{\includegraphics[width=0.49\linewidth]{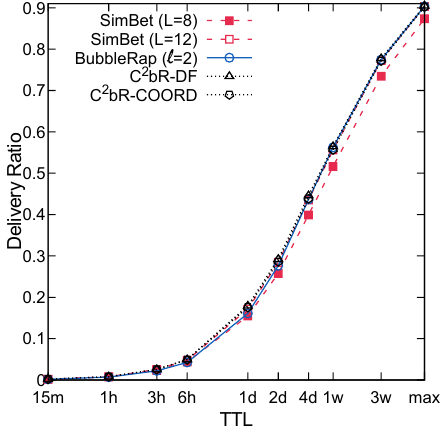}\label{DRvsTTL}}
	\subfigure[]{\includegraphics[width=0.48\linewidth]{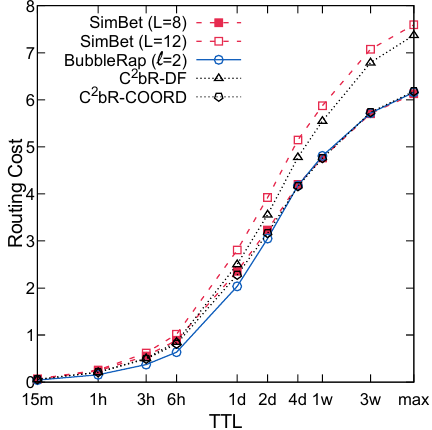}\label{OHvsTTL}}%
	\vspace{-8pt}
	\caption{Performance of C$^2$bR-DF and C$^2$bR-COORD compared to SimBet and BubbleRap vs TTL: a) delivery ratio, b) routing cost. Max three (3) hops for SimBet and BubbleRap.}
	\vspace{-18pt}
	\label{C2bR-SimBet-Bubble-2}
\end{figure}
\begin{figure*}
	\vspace{-8pt}
	\centering
	\addtolength{\subfigcapskip}{-4pt}
	\subfigure[]{\includegraphics[width=0.31\linewidth]{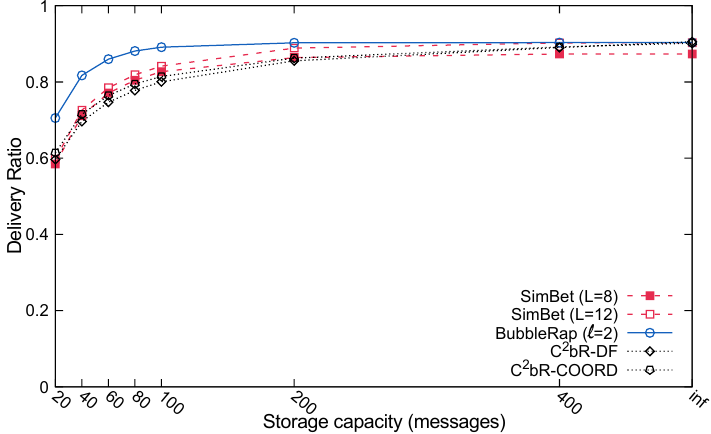}\label{DRvsBUFF}}
	\subfigure[]{\includegraphics[width=0.31\linewidth]{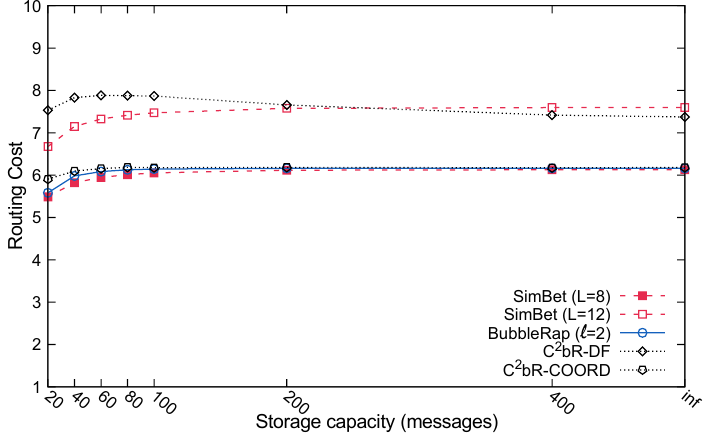}\label{OHvsBUFF}}		
	\subfigure[]{\includegraphics[width=0.31\linewidth]{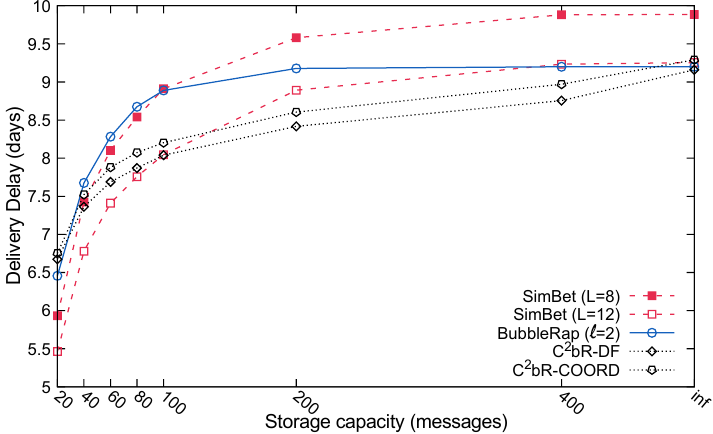}\label{DelayvsBUFF}}%
	\vspace{-8pt}
	\caption{Performance of C$^2$bR-DF and C$^2$bR-COORD compared to SimBet and BubbleRap under limited storage: a) delivery ratio, b) routing cost, c) average delay. Max three (3) hops for SimBet and BubbleRap.}
	\label{C2bR-SimBet-Bubble-3}
\end{figure*}
As a next step, additionally to TTL (again set to 20\% of trace duration), we tested a more effective rule for stopping replication which is to limit the hops that a packet can travel (\textit{hop limit} rule). The combination of this rule with the replica limit imposed by the Spray-based approach used in SimBet produced reasonable performances in terms of cost. On the contrary, the hop limit rule proved to be insufficient for delivering a reasonable performance when used in BubbleRap. Based on this finding and in order to have a fair comparison with SimBet, we also limit the number of copies in the case of BubbleRap. Since the latter is a dynamic replication algorithm, the only realistic way to do this is to limit the number of copies ($\ell$) that each node is allowed to produce. We should stress that, in the following comparison, \textit{we do not use the hop limit rule nor we limit the number of copies for C$^2$bR schemes}. We will show later that one reason for this decision is that, performance-wise, this was not necessary. The second reason pertains to the nature of C$^2$bR schemes. Discovering clusters depends on the exchange of copies and then those clusters are used to confine replication. Using other means to limit replication may damage the process of cluster formation and therefore may be harmful overall. Finally, in the following comparison we do not use C$^2$bR-CnR since the other two C$^2$bR protocols produce much better performance results.

Fig.~\ref{DRvsOHvsDelay-Multi-HL3} presents the routing cost with respect to the delivery ratio for C$^2$bR-DF and C$^2$bR-COORD. The graph also illustrates the performance of SimBet and BubbleRap for different values of $L$ and $\ell$ respectively in the case that we allow packets to travel at maximum 3 hops. A first important observation is the robustness of C$^2$bR schemes regarding their ability to confine replication, especially after the delivery of the message. Even without imposing any predetermined limit on the number of copies or on the number of hops, the cost for both algorithms presents a minimal increase compared to the previous experiment (Fig.~\ref{C2bR-SimBet-Bubble-0}). Overall, C$^2$bR-COORD strikes the best performance trade-off (the same as BubbleRap with $\ell\!\!=\!\!2$). Improving the delivery rate by $\sim\!\!0.5\%$ (SimBet $L\!\!=\!\!16$), $\sim\!\!3.5\%$ (BubbleRap $\ell\!\!=\!\!3$) or $\sim\!\!4.5\%$ (BubbleRap $\ell\!\!=\!\!4$) requires a cost that is $\sim\!\!35\%$, $\sim\!\!77.5\%$ and $\sim\!\!174\%$ greater than that of C$^2$bR-COORD, respectively. Impressively, C$^2$bR-COORD does not require any special customization. On the contrary, to achieve the best performance of BubbleRap one is required, besides customizing the community detection algorithm, to also determine the appropriate value of $\ell$. This is not straightforward because the best value depends on the connectivity properties of the network that are not known beforehand. It is evident that failure to properly set $\ell$ results in either a noticeable cutback in delivery efficiency or in a significant increase in cost (Fig~\ref{DRvsOHvsDelay-Multi-HL3}). Furthermore, it is critical for BubbleRap to choose the proper hop limit. Fig~\ref{DRvsOHvsDelay-Multi-HL4} is similar to fig.~\ref{DRvsOHvsDelay-Multi-HL3} but for a limit of 4 hops for SimBet and Bubble Rap. Clearly, increasing the hop limit destroys the performance trade-off for BubbleRap regardless of $\ell$. This illustrates the importance of yet another parameter that requires non-trivial customization because it depends on the network properties which may not be known, especially at setup time. Fig.~\ref{C2bR-SimBet-Bubble-1} implies that a similar customization problem also applies to SimBet, i.e., a misfire in customization of either $L$ or the hop limit significantly affects its performance. On the other hand, C$^2$bR schemes do not depend on any similar customization and although C$^2$bR-COORD outperforms C$^2$bR-DF, the latter is very close and its performance is competitive to those of SimBet and BubbleRap.

Another interesting finding is that C$^2$bR schemes perform efficiently compared to the other algorithms under different values of packet TTL. In other words, the rules for stopping replication in C$^2$bR schemes do not impair the ability of the cluster-based approach to timely deliver packets. Fig.~\ref{C2bR-SimBet-Bubble-2} presents the performance of all algorithms for different values of TTL from a minimum of 15 mins to a maximum that equals the 20\% of the Reality trace duration. For SimBet and BubbleRap we present the best performances, i.e., $L\!\!=\!\!8$ and $L\!\!=\!\!12$ with a limit of 3 hops for SimBet and $\ell\!\!=\!\!2$ with a limit of 3 hops for BubbleRap. Both C$^2$R-DF and C$^2$R-COORD achieve delivery performances similar to BubbleRap and SimBet with $L\!\!=\!\!12$ for all TTL values (Fig.~\ref{DRvsTTL}). In fact, both the C$^2$bR schemes slightly outperform the other two for medium TTL values. Only SimBet with $L\!\!=\!\!8$ lags significantly in delivery efficiency which comes as a trade-off for reducing cost (Fig.~\ref{OHvsTTL}). C$^2$bR-DF outperforms SimBet with $L\!\!=\!\!12$ and C$^2$bR-COORD performs similar to BubbleRap ($\ell\!\!=\!\!2$) in terms of cost although, contrary to their counterparts, they do not require any network-dependent customization.

As a final test, we explored the performance of the algorithms in the case of limited node storage (Fig.~\ref{C2bR-SimBet-Bubble-3}). Reasonably, the delivery efficiency of all algorithms declines as the available storage gets smaller. Both C$^2$bR-DF and C$^2$bR-COORD achieve performances that are competitive to those of SimBet and BubbleRap. This is especially true if we keep in mind that the presented versions of SimBet and BubbleRap are the ones with the optimal replication lever for each algorithm. This is critical since controlling replication allows more free storage space and therefore minimizes the packet drop rate. Needless to say that producing the optimal replication level for SimBet and BubbleRap calls for a network-dependent fine-tuning which may not even be possible in a real-life setting. As a last remark, BubbleRap exhibits an increased resilience to limited storage. This is mainly due to the method we implemented for controlling replicas which inherently imposes load balancing since each node is allowed to create the same number of copies.

\section{Conclusion}\label{sec-concl}
Despite their flexibility in effectively operate in delay-tolerant networks with diverse characteristics, dynamic replication schemes are inclined towards over-replication. To deal with the problem, we first made the observation that the utility values observed by a node through its contacts form clusters. We validated that these clusters can be identified by a node using lightweight clustering algorithms. Then, we delineated a novel forwarding policy that can be used to transform the decision making process of traditional dynamic replication schemes to one that relies on cluster-based decisions. More specifically, the key concept in our approach to forward a packet through clusters of increasing delivery capability, in contrast to the existing approach which is to create replicas in nodes of increasing utilities. We also extended our cluster-based approach to work with two utility functions at the same time. This extension is tailored for routing in mobile social networks. We experimentally demonstrated the significant performance benefits of cluster-based replication when operating either with one or two utility functions. We also validated that our approach is robust in a set of networks with diverse characteristics without the need for a complex and non-trivial pre-configuration.

\appendix
\setcounter{equation}{0}
\setcounter{figure}{0}

\section{Performance with limited node storage}\label{limited-storage-experiment}

\begin{figure*}[b]
	\vspace{-16pt}
	\centering
	\addtolength{\subfigcapskip}{-4pt}
	\subfigure[]{\includegraphics[width=0.32\linewidth]{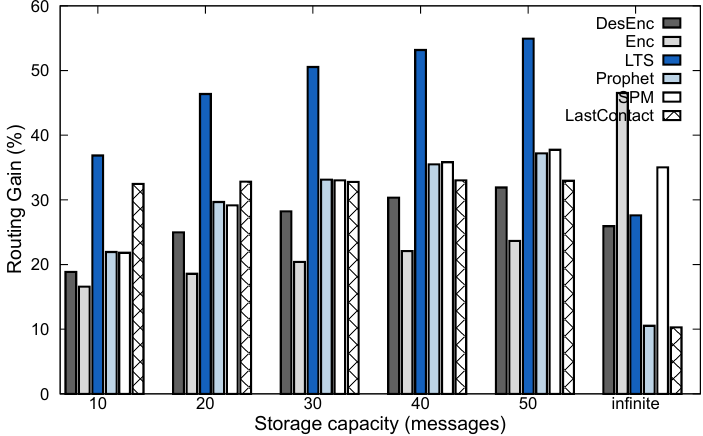}\label{CnRKMeansOHCampBufReality}}\hspace{-2pt}
	\subfigure[]{\includegraphics[width=0.32\linewidth]{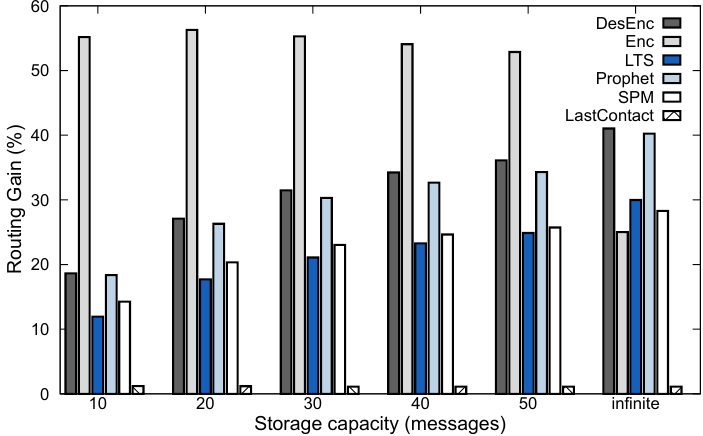}\label{DFKMeansOHCampBufReality}}\hspace{-2pt}		
	\subfigure[]{\includegraphics[width=0.32\linewidth]{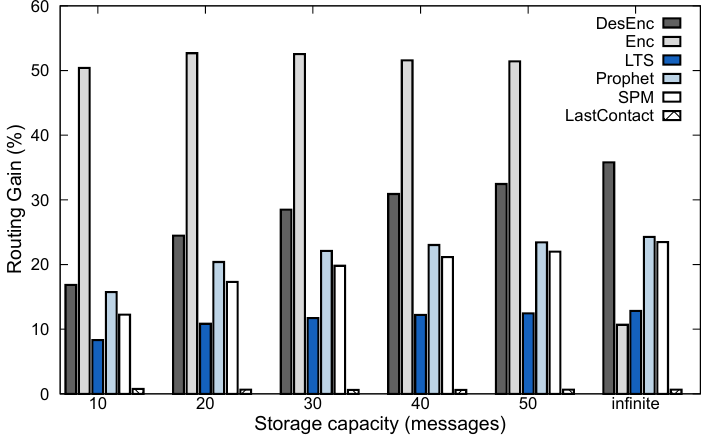}\label{COORDKMeansOHCampBufReality}}\hspace{-2pt}
	\subfigure[]{\includegraphics[width=0.32\linewidth]{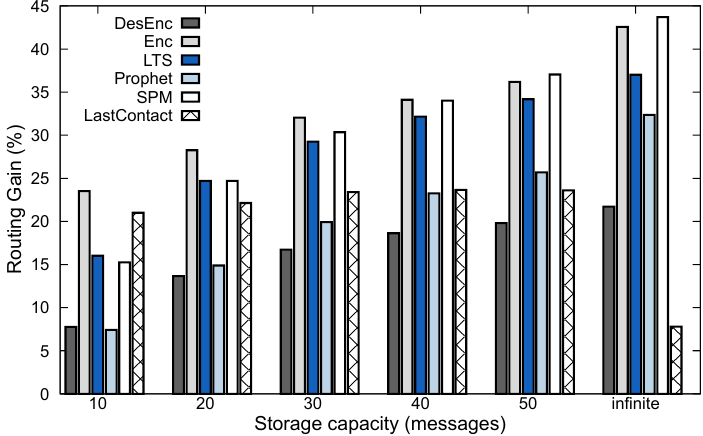}\label{CnRKMeansOHCampBufCambridge}}\hspace{-2pt}
	\subfigure[]{\includegraphics[width=0.32\linewidth]{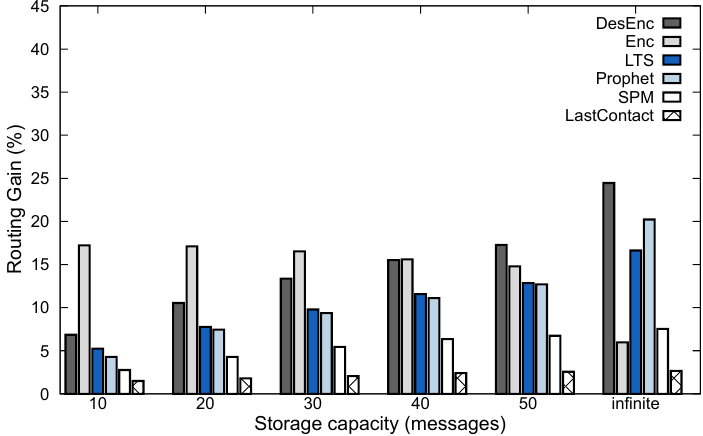}\label{DFKMeansOHCampBufCambridge}}\hspace{-2pt}		
	\subfigure[]{\includegraphics[width=0.32\linewidth]{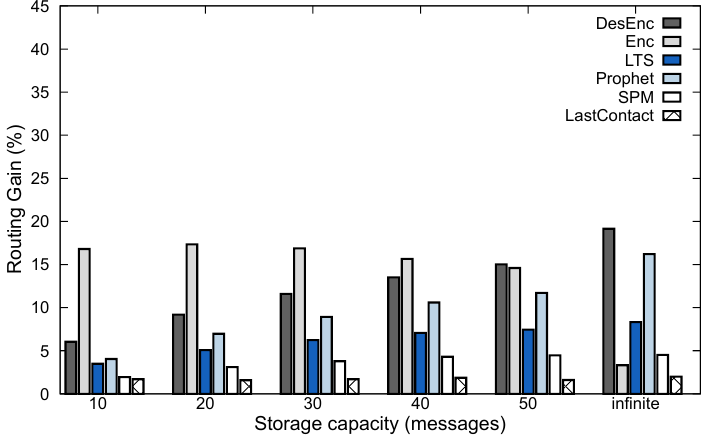}\label{COORDKMeansOHCampBufCambridge}}\hspace{-2pt}%
	\vspace{-9pt}
	\caption{Routing gain of the CbR approach vs node's storage capacity for various utility metrics in the Reality trace [(a) CbR-CnR, (b) CbR-DF, and (c) CbR-COORD] and in the Cambridge trace [(d) CbR-CnR, (e) CbR-DF, and (f) CbR-COORD].}
	\vspace{-6pt}
	\label{OH-vsBUF-results}
\end{figure*}
\begin{figure*}
	\centering
	\addtolength{\subfigcapskip}{-4pt}
	\subfigure[]{\includegraphics[width=0.49\linewidth]{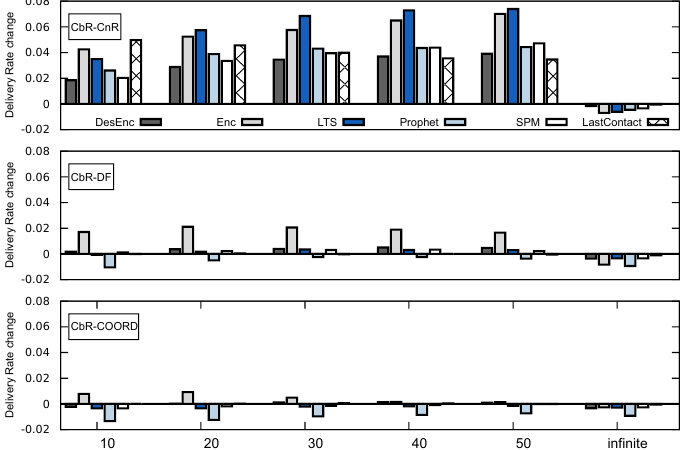}\label{CnRKMeansDRCampBufReality}}\hspace{-2pt}	
	\subfigure[]{\includegraphics[width=0.49\linewidth]{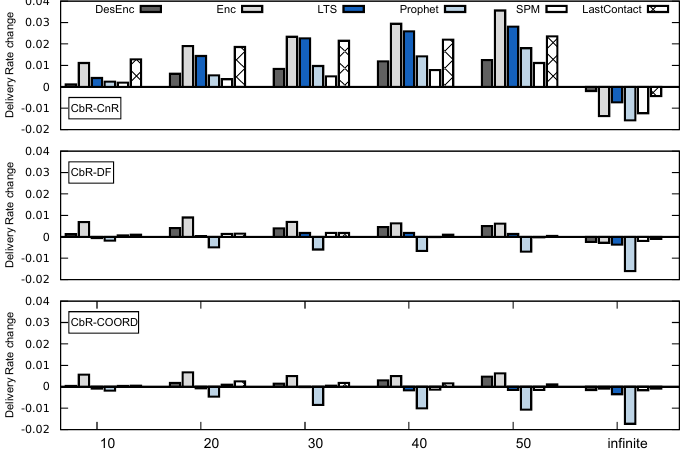}\label{COORDKMeansDRCampBufCambridge}}\hspace{-2pt}%
	\vspace{-9pt}
	\caption{Delivery rate change (when implementing CbR) vs storage capacity for various utility metrics in: a) the Reality trace, and b) the Cambridge trace}
	\label{DR-vsBUF-results}
\end{figure*}
In this experiment we focus on the Reality and Cambridge traces and examine the case of limited storage, i.e., a node can only store a limited number of packets. More specifically, we test the performance of CbR with respect to the storage capacity of nodes. Fig.~\ref{OH-vsBUF-results} illustrates the routing gain when CbR is used for both traces. Fig.~\ref{CnRKMeansDRCampBufReality} presents the delivery rate change for CbR-CnR, CbR-DF and CbR-COORD for various utility functions in the Reality trace while Fig.~\ref{COORDKMeansDRCampBufCambridge} presents the same delivery rate change in the Cambridge trace. Clearly, when storage is limited, all CbR flavors provide not only significantly better routing cost gains but also better delivery efficiency compared to the unlimited storage case. We found this to be true not only for the two presented but also for the rest of the traces. This is reasonable since reducing the routing load significantly alleviates congestion and cuts down the packet drop rate. This is also why the improvement is bigger for CbR-CnR since in this case congestion is more severe. On the other hand, for CnR-DF and CbR-COORD the improvement, although evident, is limited because both DF and COORD are able to effectively reduce transmissions, and thus congestion, on their own. Overall, under limited storage, CbR-CnR combines improvements in both the routing cost and the delivery efficiency compared to CnR. At the same time, CbR-DF and CbR-COORD provide significant routing gains and a delivery performance which is slightly better or similar to their baseline algorithms, i.e., DF and COORD respectively.

Looking in more detail in the routing cost performance, CbR outperforms by a wide margin the baseline algorithm (positive routing gain) in all cases, i.e., combinations of trace and utility function. The only exception is the LastContact utility when used with CbR-DF and CbR-COORD where the gain is minimal. This can be associated with the structure of this utility, i.e., a destination independent utility that produces values with little diversity, especially in traces with sparse contacts. For the rest of the cases, reasonably the general trend is that the gain increases with the storage capacity while it is still significant for very small buffer sizes. This is because less packets are dropped and this provides more opportunities for pruning replicas. However, there are two exceptions where the gain decreases. The first is the case of infinite buffer size in CbR-CnR in the Reality dataset (Fig.~\ref{CnRKMeansOHCampBufReality}). To shed some light, observe that implementing CbR on top of CnR significantly improves the delivery efficiency when the storage capacity is limited (Fig.~\ref{CnRKMeansDRCampBufReality}). This improves CbR's routing cost $T_{CbR}$ because the latter is normalized to the number of delivered packets. As a result, the routing gain $1\!-\!T_{CbR}/T$ appears to increase when the storage is limited compared to the case of unlimited storage. Note that the phenomenon does not appear to this extent in the Cambridge trace because the increase in delivery efficiency is smaller. Moreover, the utilities that tend to over-replicate, such as Enc, are unaffected by this phenomenon. This is because over-replication is more severe when no storage limitation exists and the effectiveness of CbR in reducing the routing gain dramatically increases in conditions of over-replication. The second exception to the gain increasing trend is observed when Enc is used in CbR-DF (Fig.~\ref{DFKMeansOHCampBufReality} and~\ref{DFKMeansOHCampBufCambridge}) and CbR-COORD (Fig.~\ref{COORDKMeansOHCampBufReality} and\ref{COORDKMeansOHCampBufCambridge}). 
This can be explained with the same reasoning discussed for the previous exception with the additional note that, contrary to the case of CbR-CnR with Enc, here over-replication is limited due to CbR-DF and CbR-COORD.

\setcounter{equation}{0}
\setcounter{figure}{0}

\section{Updating the Clustering Result}\label{appendix-sec-updating}
\begin{figure*}[b]
	\centering
	\addtolength{\subfigcapskip}{-4pt}
	\subfigure[]{\includegraphics[width=0.29\linewidth]{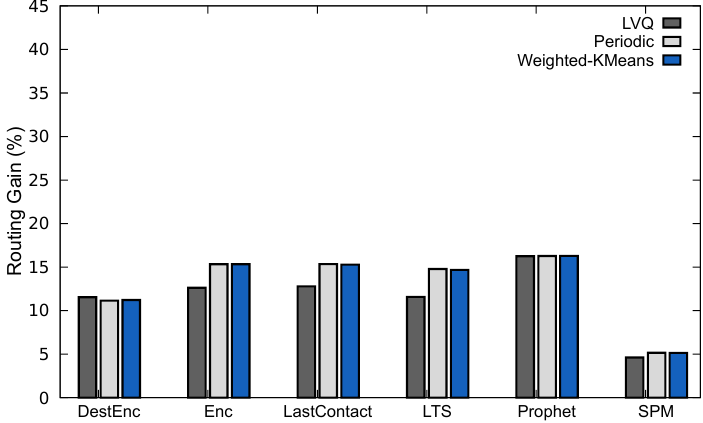}\label{CbR-DF-OH_Updt_Milano}}\hspace{-2pt}
	\subfigure[]{\includegraphics[width=0.29\linewidth]{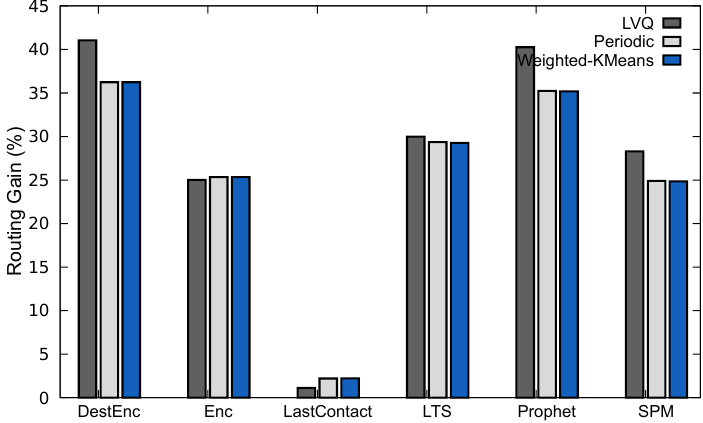}\label{CbR-DF-OH_Updt_Reality}}\hspace{-2pt}
	\subfigure[]{\includegraphics[width=0.29\linewidth]{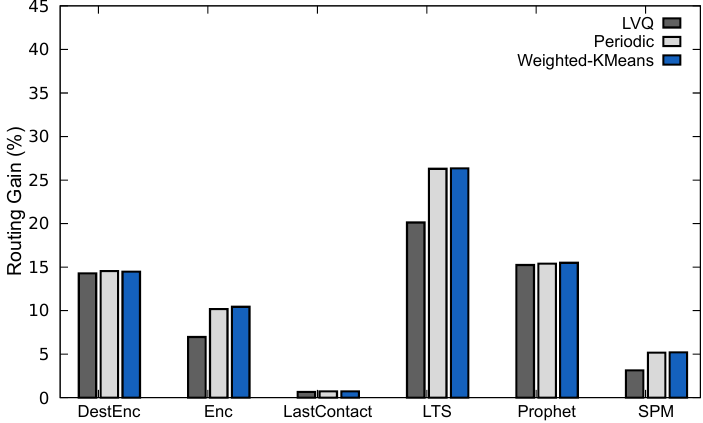}\label{CbR-DF-OH_Updt_Infocom05}}\hspace{-2pt}
	\subfigure[]{\includegraphics[width=0.29\linewidth]{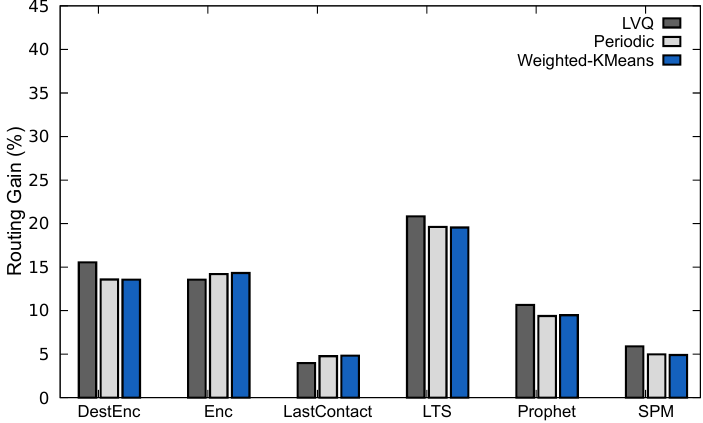}\label{CbR-DF-OH_Updt_Sigcomm}}\hspace{-2pt}		
	\subfigure[]{\includegraphics[width=0.29\linewidth]{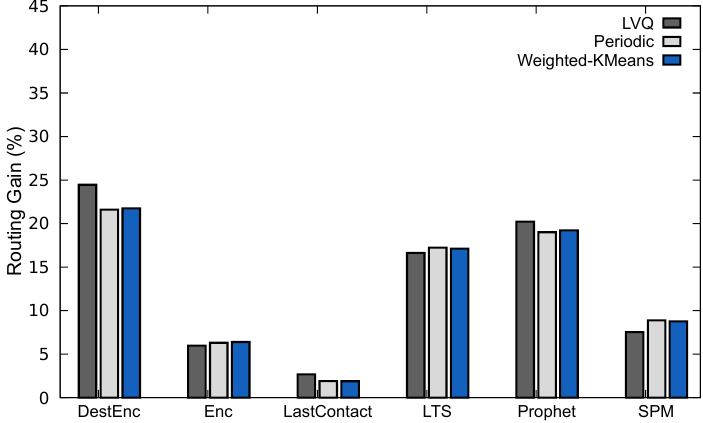}\label{CbR-CnR-OH_Updt_Cambridge}}\hspace{-2pt}		
	\subfigure[]{\includegraphics[width=0.29\linewidth]{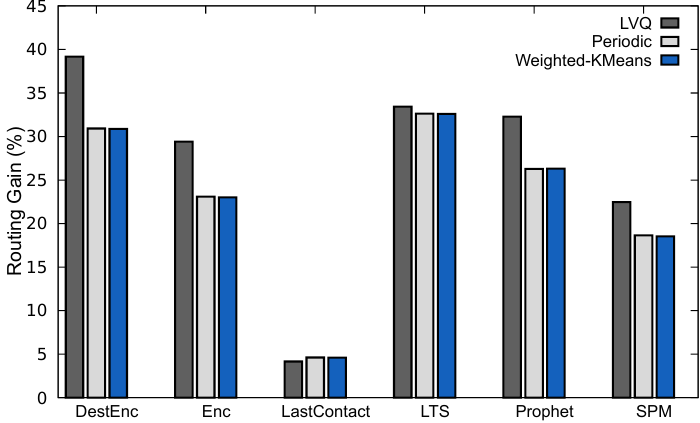}\label{CbR-CnR-OH_Updt_NCCU}}%
	\vspace{-12pt}
	\caption{Routing gain of three different updating methods when implemented in CbR-DF for various utility metrics in: a) Milano, b) Reality, c) Infocom05, d) Sigcomm, e) Cambridge, and f) NCCU traces}
	\vspace{-12pt}
	\label{UpdatingMethods-results}
\end{figure*}

In Section~\ref{subsec-updating} we discussed the requirement for refreshing the utility values recorded by a node and accordingly its clustering result. Since we observed rather simple and smooth changes in the recorded data, we opted for LVQ as the refreshing function due to its low complexity. However, alternative update methods could be examined. Here, we evaluate the performance of two alternative update methods and justify our choice of using LVQ. 

The first method is \textit{periodic $k$-Means}, i.e., a periodic, window-based execution of $k$-Means algorithm. More specifically, after completing the training period and detecting utility clusters, a node continues to record new utility values. After collecting $T_{P}$ new samples, the node re-evaluates the utility clusters using the $k$-Means algorithm and the $W$ most recently recorded utility values. We call $T_{P}$ the update period and $W$ the update window. This method is useful for operating CbR in conditions that the optimal size of the training period ($N_{TR}$) may vary. In such a case, one should select $N_{TR}=T_{P}$ and set $T_{P}$ to a relatively small value. On the other hand, $W$ should be set to a relatively large value. With this setting, $T_{P}$ serves as the minimum training period while $W$ as the maximum one. A first clustering result will be available after $T_{P}$ samples are collected. Then, the clustering result will be updated every $T_{P}$ samples to eventually include up to the $W$ more recent samples and secure the correct detection of clusters. Until reaching this point, clearly the clustering result will be less accurate if a long training period is required, i.e., the actual $N_{TR}\rightarrow W$. However the algorithm can still operate since, as we discussed in Section V, even in this case there is no way to produce over-replication compared to the baseline algorithm.

The second update approach extents the periodic one by using a concept known as \textit{weighted $k$-Means}~\cite{weighted-kmeans}. The idea here is to assign to each recorded utility value a weighting factor and then execute the $k$-Means algorithm. The weight for each recorded value decreases with the age of this value, i.e., an older recorded value is assigned a smaller weight, thus providing a node with the ability to adjust its clustering result to more recent utility values. More specifically, we assign to each recorded utility value $u$ a weight:
\begin{equation}
	w(u) = e^{-i/R}
\end{equation} 
where $i$ is the index of $u$ if all utilities in the window $W$ are ordered by the recording time and $R$ is a constant that controls the weight decaying rate. The objective function is now:
\begin{equation}
	\sum_{i=1}^{k}\sum_{u \in C_{i}} w(u)||u-c_{i}||^{2}
\end{equation}
where $C_{1},C_{2},\ldots,C_{k}$ are the $k$ clusters of utility values. The value $c_{i}$ is the mean of $C_{i}$:
\begin{equation}
	c_{i} = \frac{1}{\sum\limits_{u \in C_{i}}w(u)}\sum_{u \in C_{i}} uw(u)
\end{equation}
Observe that the traditional $k$-Means algorithm can be seen as a special case where $w(u)$ is constant, i.e. $w(u)\!=\!w, \forall u$.

We implemented both updating methods in Adyton~\cite{Adyton} and assessed their performance compared to LVQ. Similar to the traditional implementation of $k$-Means, here, we also use the Silhouette criterion to automatically select the best value of $k$. We assumed unlimited storage at nodes in order to avoid other interfering factors. For periodic $k$-Means, we report results for $T_{P}\!=\!50$ and $W\!=\!50$. Although we also tested various combinations of values for $T_{P}$ and $W$, we witnessed insignificant performance variations. Regarding weighted $k$-Means, we report the results for the same values of $T_{P}$ and $W$ and for $R\!=\!400$. Again, when using different values of $R$, we observed only slight performance variations. We used CbR-DF as the reference algorithm and produced three algorithm versions corresponding to the three updating methods. Then, we captured the performance of those three versions using various utility functions in all the investigated traces. Fig.~\ref{UpdatingMethods-results} illustrates the routing gain of the three CbR-DF versions compared to the simple DF algorithm. For the majority of utility functions and trace combinations the three schemes achieve similar performance. The result is reasonable since we have observed that the number of utility clusters that a node detects rarely changes. Instead, the time evolution mostly concerns the center of the clusters, a type of evolution that LVQ can handle as efficiently as the other two update methods. In fact, for some utility functions such as DestEnc, PRoPHET, SPM and LTS, LVQ performs consistently and noticeably better than the other two schemes, an indication of the smooth adaptation to the changing network conditions. On the other hand, LVQ is slightly lagging in most cases when a destination independent (DI) utility is used, e.g., Enc and LastContact. Recall that a DI utility $U_{v}$ aims to capture the generic importance of $v$, therefore it is built based on contact information regarding multiple possible destination nodes instead of a single one in the case of destination dependent (DD) utilities. This makes a DI utility a more mutable quantity compared to a DD one. In any case, the routing gain lag of LVQ is extremely limited and can be considered an acceptable trade-off for its lower computational complexity.
The same picture of minimal performance variations between the three schemes also appears when examining the normalized delivery rate and delay.
Regarding the delivery rate change, besides an $\sim\!\!1\%$ improvement in favor of LVQ when LTS is used, we found that the three schemes yield practically the same performance (maximum variation $0.4\%$). The same observation applies to the delay change where the maximum variation was $1.1\%$.

\setcounter{equation}{0}
\setcounter{figure}{0}

\section{Performance and Complexity Considerations}\label{subsec-thperformance}

Since $CbR$ is not a standalone algorithm but it rather operates on top of existing dynamic replication algorithms, its worst-case replication performance coincides with the one of the underlying algorithm. Let us consider the case that $CbR$ is not implemented and let $v$ be a packet carrier while $T_{v}$ is the set of $v$'s encounters that receive a replica from $v$, ordered by the time of arrival. Let also $T'_{v}$ be the same set when $CbR$ is implemented. Moreover, when $CbR$ is used, assume that $v$ identifies clusters $c_{1},c_{2},\ldots,c_{k}, \, k>1$. Observe in Fig.~\ref{pseudocodeCbR-Replication} that in $CbR$ node $v$ implements the replication strategy of the underlying algorithm until it encounters a node from a better ranked cluster. In this case, $CbR$ stops replication within the same cluster and creates at most one replica for every better ranked cluster. In other words, if $v$ belongs to cluster $c_{i}$ then
\begin{equation}\label{basicres}
	|T'_{v}| \leq |T_{v}^{C}| + min\{|T_{v}|-|T_{v}^{C}|, k-i\}
\end{equation}
where $|T_{v}^{C}|$ is the number of the first consecutive replicas created to nodes in the same cluster ($|T_{v}^{C}| \leq |T_{v}|$) and $k-i$ is the number of clusters with a rank higher than of $c_{i}$.

Let us denote $p_{c}$ the probability that $v$ encounters a candidate packet carrier (i.e., with a suitable utility) that belongs to the same cluster ($c_{i}$). Let also $P_{m}$ be the probability that exactly $m$ consecutive replicas are created to nodes in $c_{i}$ before creating a replica outside of that cluster. Since replication within $c_{i}$ stops when a replica is created outside the cluster, $P_{m}$ is the probability that $v$ performs the first $m$ consecutive replications to nodes belonging in $c_{i}$ (probability $p_{c}^{m}$) while the $(m+1)$-th replication is to a node outside $c_{i}$ (probability $1-p_{c}$). Therefore, $P_{m} = p_{c}^{m}(1-p_{c}), \forall m < |T_{v}|$. Note that this is an upper limit for $P_{m}$ since $p_{c}$ may decrease as $m$ increases. This is because in DF and COORD each replication increases the threshold therefore it is more difficult to find a node in the same cluster and with a higher utility. 
Clearly, $P_{|T_{v}|}$, i.e., the probability of creating $i=|T_{v}|$ replicas in the same cluster, is $p_{c}^{|T_{v}|}$. Consequently, it can be shown that
\begin{equation}\label{firstres}
	E(|T_{v}^{C}|)=\frac{p_{c}-p_{c}^{|T_{v}+1|}}{1-p_{c}}
\end{equation}
\noindent i.e., on average $E(|T_{v}^{C}|)$ out of the total $|T_{v}|$ replicas will be created within the same cluster before replicating to a higher ranked cluster. Thus, using (\ref{basicres}) we find that
\begin{equation}\label{firstsecondres}
	E(|T'_{v}|) \leq \frac{p_{c}\!-\!p_{c}^{|T_{v}+1|}}{1-p_{c}}\!+\!min\{|T_{v}|\!-\!\frac{p_{c}\!-\!p_{c}^{|T_{v}+1|}}{1-p_{c}}, k\!-\!i\}
\end{equation}
\noindent Note that when $p_{c}\!=\!1$, i.e., when $v$ encounters only nodes in the same cluster, then $E(|T_{v}^{C}|)\!=\!|T_{v}|$ and $E(|T'_{v}|)\!=\!|T_{v}|$, i.e., node $v$ performs exactly the same number of replications even if $CbR$ is used. This happens when the clustering of utility values is not possible or when $v$ is the only node in the cluster, i.e., $|c_{i}|\!=\!1$. In both cases, $CbR$ does not create any overhead but replicates the performance of the underlying algorithm (see Fig.~\ref{pseudocodeCbR-Replication}). Furthermore, it is possible that $E(|T'_{v}|)\!=\!|T_{v}|$ even if a reasonable clustering exists. This is the case that $v$ belongs to $c_{k}$, i.e., the highest ranked cell, because in this case there are no nodes with higher utility that are outside cluster $c_{k}$, therefore $p_{c}\!=\!1$. However, when $v$ is in any other cluster (i.e., $p_{c} < 1$), we can show from (\ref{firstres}) that $E(|T'_{v}|) < |T_{v}|, \,\, \forall \, |T_{v}|$, i.e., $CbR$ always performs better than the underlying algorithm. In networks where a non-trivial utility clustering is present we expect that $p_{c}\!\propto\!1/k$, therefore according to (\ref{firstres}) the benefits of using $CbR$ can grow large even when a relatively small number of clusters exists. Note that when $p_{c} < 0.5$, i.e., even for relatively large values of $p_{c}$, according to (\ref{firstres}), $E(|T_{v}^{c}|) = \delta < 1, \;\; \forall \; |T_{v}|$. As a consequence, (\ref{firstsecondres}) results in 
\begin{equation}\label{secondres}
	E(|T'_{v}|) \leq \delta +  min\{(1-p_{c})|T_{v}|, k-i\}
\end{equation}
Since $k$ is typically rather small and $k-i$ is even smaller, the second term in (\ref{secondres}) equals $k-i$ in large networks where $|T_{v}|$ is also large. Therefore, in such networks $E(|T'_{v}|)$ is $O(k)$, i.e., when $CbR$ is used, the average number of replicas (or equivalently transmissions) performed by a packet carrier does not depend on the network size $N$ but on the number of clusters that the node has identified. For non-trivial clustering results $k$ is typically small and also $k \ll N$.

Regarding the cost for implementing $CbR$, one can identify two components; one related to the exchange of the appropriate information between encountering nodes (communication cost) and the other related to the cost of executing the $k$-Means algorithm. As discussed earlier, $CbR$ does not involve any additional communication cost compared to the underlying algorithm. No other information, besides that exchanged by the underlying algorithm, is required for a node to train and calculate the utility clusters. In what concerns the $k$-Means algorithm, its computational complexity is $O(N_{TR}k)$, i.e., $k$-Means is linear both with respect to the number of values during a training period ($N_{TR}$) and to the number of clusters ($k$). We will show in the following that, typically, $N_{TR} \ll N$, where $N$ is the number of nodes. Furthermore, we will also show that, in the wide range of evaluated real-life networks, $k$ is always small. But what is of the highest importance is the fact that the cost of $k$-Means algorithm is an one-time cost. This is because the algorithm is executed only once at the end of the training period. Afterwards, any new utility value is integrated into the existing clustering result using the LVQ algorithm. The running time for this process is clearly $O(k)$. Keeping in mind that state of the art processors can provide several DMIPS (Dhrystone Million Instructions Per Second)~\cite{arm-processors} per mW~\cite{embedded-software}, it is clear that not only it is possible for a mobile node to implement CbR in a real-time scenario but also that the energy cost of finding and updating clusters is clearly negligible. Furthermore, researchers have provided evidence that the energy consumption in a wireless interface is hugely higher than the one related to processing~\cite{smartphone-analysis,energy-greencom}. Consequently, the ability of $CbR$ to reduce transmissions dominates the energy balance and manifests its energy efficiency.

\end{document}